\documentclass[twocolumn]{aastex62}

\usepackage{natbib}
\usepackage{float}
\graphicspath{{./}{figures/}}

\accepted{August 2023}
\submitjournal{ApJ}

\shorttitle{Kepler's SNR Ejecta Mass Ratios}
\shortauthors{Holland-Ashford et. al.}

\begin{document}

\title{Estimating Ejecta Mass Ratios in Kepler's SNR: \\
Global X-Ray Spectral Analysis Including {\it Suzaku} Systematics and Emitting Volume Uncertainties}

\correspondingauthor{Tyler Holland-Ashford}
\email{tyler.holland-ashford@cfa.harvard.edu}
\author{Tyler Holland-Ashford}
\affil{Center for Astrophysics $|$ Harvard \& Smithsonian, 60 Garden St, Cambridge MA 02138, USA}
\author{Patrick Slane}
\affil{Center for Astrophysics $|$ Harvard \& Smithsonian, 60 Garden St, Cambridge MA 02138, USA}

\author{Laura A. Lopez}
\affil{Department of Astronomy, The Ohio State University, 140 W. 18th Ave., Columbus, Ohio 43210, USA}
\affil{Center for Cosmology and AstroParticle Physics, The Ohio State University, 191 W. Woodruff Ave., Columbus, OH 43210, USA}
\affil{Flatiron Institute, Center for Computational Astrophysics, NY 10010, USA}

\author{Katie Auchettl}

\affil{OzGrav, School of Physics, The University of Melbourne, Parkville, Victoria 3010, Australia}
\affil{ARC Centre of Excellence for All Sky Astrophysics in 3 Dimensions (ASTRO 3D)}
\affil{Department of Astronomy and Astrophysics, University of California, Santa Cruz, CA 95064, USA}

\author{Vinay Kashyap}
\affil{Center for Astrophysics $|$ Harvard \& Smithsonian, 60 Garden St, Cambridge MA 02138, USA}

\date{August 2023}

\begin{abstract}
The exact origins of many Type Ia supernovae---progenitor scenarios and explosive mechanisms---remain uncertain. In this work, we analyze the global {\it Suzaku} X-Ray spectrum of Kepler's supernova remnant in order to constrain mass ratios of various ejecta species synthesized during explosion. Critically, we account for the {\it Suzaku} telescope effective area calibration uncertainties of 5--20\% by generating 100 mock effective area curves and using Markov Chain Monte Carlo based spectral fitting to produce 100 sets of best-fit parameter values. Additionally, we characterize the uncertainties from assumptions made about the emitting volumes of each model plasma component and find that these uncertainties can be the dominant source of error. We then compare our calculated mass ratios to previous observational studies of Kepler's SNR and to the predictions of Ia simulations. Our mass ratio estimates require a $\sim$90\% attenuated $^{12}$C$+^{16}$O reaction rate and are potentially consistent with both near- and sub-M$_{\rm Ch}$ progenitors, but are inconsistent with the dynamically stable double detonation origin scenario and only marginally consistent with the dynamically unstable dynamically-driven double-degenerate double detonation (D$^6$) scenario.

\end{abstract}

\keywords{ISM: supernova remnants -- methods: data analysis -- supernovae: individual (Kepler's SNR) -- techniques: imaging spectroscopy -- X-Rays: ISM -- instrumentation: X-Ray telescopes}

\section{Introduction}

Type Ia supernovae (SNe) are the thermonuclear explosions of carbon-oxygen white dwarfs (CO WDs) resulting from binary interactions with a companion \citep{hoyle60, nomoto82}. These energetic events play a critical role in cosmic nucleosynthesis, forming many of the heavy elements that are incorporated into new generations of stars \citep{truran67,matteucci89, kobayashi06}. However, the exact processes that lead to SNe Ia are uncertain. 

Historically, Type Ia models have been split into single-degenerate (SD) and double-degenerate (DD) origins, referring to the primary WD having a non-WD vs WD companion \citep{whelan73, iben84}. These scenarios are often linked with the mass of the primary WD; SD explosions typically involve the primary WD accreting material until it exceeds the Chandrasekhar limit---M$_{\rm Ch}\sim$1.4M$_\odot$---and explodes \citep{colgate66}. DD explosions are classically explained as the in-spiral and merging of two sub-M$_{\rm Ch}$ WDs due to gravitational wave emissions \citep{seitenzahl17}.

More recently, other scenarios have shown promise in matching observed SNIa properties. In the double detonation (DDet) scenario, stable accretion of He builds up on the surface of a WD and eventually ignites, inducing an explosion in even sub-M$_{\rm Ch}$ WDs \citep{taam80, livne90, alan19}. In the Dynamically-Driven Double-Degenerate Double Detonation (D$^6$) variant, the He accretion is unstable and leads to an earlier ignition of the outer He layer and a possible surviving companion WD \citep{guillochon10,dan11,shen18a}.

Models of each of these scenarios make different predictions about explosion and progenitor properties that can be connected to astrophysical observables: e.g., nucleosynthesis of intermediate-mass elements (IME: e.g., oxygen, magnesium, neon, silicon, sulfur, argon, and calcium) and iron-group elements (IGE or Fe-group; e.g., chromium, manganese, iron, nickel). The production of these different elements can differ drastically with progenitor metallicity \citep{bravo10}, the amount of carbon simmering prior to explosion \citep{chamulak08}, the central density of the progenitor which leads to different types of burning processes \citep{iwamoto99}, and more. For example, while near-M$_{\rm Ch}$ SNe Ia have historically thought to be the dominant formation processes for Mn due to their high density cores (see e.g., \citealt{seitenzahl11}), recent studies have shown that double detonations of sub-M$_{\rm Ch}$ WDs can also produce super-solar [Mn/Fe] (e.g., \citealt{lach20}).

Kepler's SNR is a young, $\sim$420-year old supernova remnant (SNR) with a Type Ia origin \citep{baade43, kinugasa99, badenes06, reynolds07}. It is one of the most well-studied SNRs, with its relatively close distance ($\sim$5~kpc; \citealt{sankrit16, ruizlapuente17}) and young age indicating that most of the emission is coming from heated-up ejecta material \citep{kinugasa99, reynolds07} rather than swept-up material from the interstellar medium (ISM). Although it seems to be interacting with dense circumstellar material (CSM; \citealt{blair07,williams12,katsuda15}) which supports a SD origin (e.g., \citealt{patnaude12}), a surviving companion that would be a smoking gun has not yet been found (e.g., \citealt{kerzendorf14}).

In this paper, we analyze X-Ray spectra of Kepler's SNR in order to compare observed ejecta mass ratios to the predictions of Type Ia SNe simulations. We perform Markov Chain Monte Carlo (MCMC) fitting to the 0.6--8.0~keV {\it Suzaku} X-Ray spectrum of the entire SNR, a process that enables us to calculate the total mass ratios of many X-Ray-emitting ejecta species. Compared to past works on Kepler's SNR, our analysis is more comprehensive, fitting the broad 0.6--8.0~keV spectrum instead of narrower bandpasses and fitting the entirety of the SNR instead of a specific region. Additionally, we are the first to quantify the effects of both telescope systematics and plasma emitting volume assumptions on estimated SNR mass ratios, ensuring that our results are as robust as possible and providing a framework for future high signal-to-noise SNR spectral analysis.

Our paper is formatted as follows: in Section~\ref{sec:methods}, we describe our data and the procedures used to analyze the data---including accounting for telescope calibration uncertainties and the effects of emitting volume assumptions. In Section~\ref{sec:results}, we present our final mass ratio estimates. In Sections~\ref{sec:pastwork} and \ref{sec:simuls}, we compare our mass ratio estimates to the results of past observational papers and to the predictions of Type Ia simulations.

\begin{figure*}
\begin{center}
\includegraphics[width=0.93\textwidth]{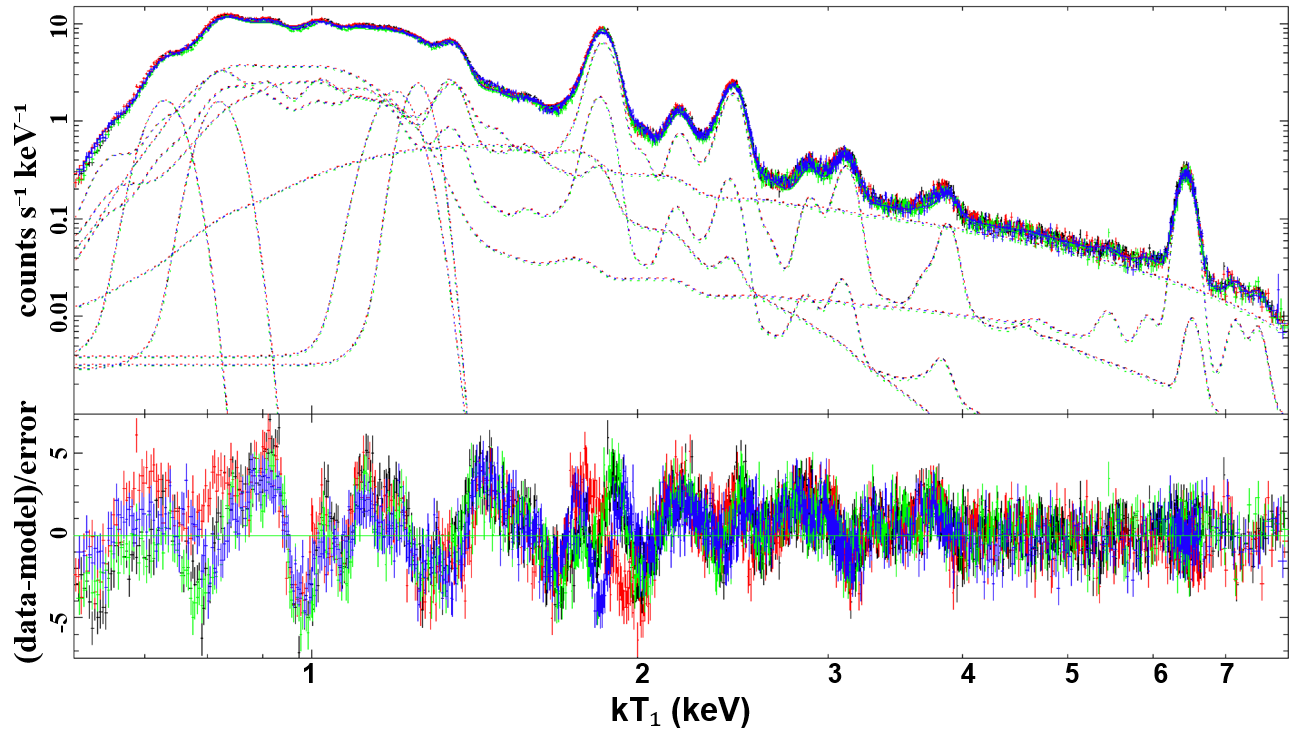}
\end{center}
\vspace{-5mm}
\caption{\footnotesize{{\it Suzaku} X-Ray spectra of Kepler's SNR and one of our multi-component spectral fits. Each color is from a different observation, the dotted lines are the individual components of the fit, and the solid black line (nearly-perfectly overlaid on the data) is the summed total fit. Although there are many regions where the fit has high residuals, the fit still broadly matches the data.
}}
\label{fig:specfit}
\end{figure*}

\section{Observations and Data Analysis}
\label{sec:methods}

\subsection{{\it Suzaku} Data Reduction}
Kepler's SNR was observed by {\it Suzaku} 8 times over 3 years and, with an angular diameter of $\sim$4\arcmin\ \citep{green19}, the SNR is fully enclosed by the field of view of {\it Suzaku}'s X-Ray Imaging Spectrometer detectors (XIS; \citealt{koyama07}). We analyze the 4 longest observations that were taken within 1~year of each other: ObsIDs 505092040, 505092070, 505092020, and 505092050 for a total of $\sim$475~ks. We only used data from the front-illuminated XIS0 CCD detector and found that using more observations or additional detectors didn't improve the fits. 

We processed the data using the HEADAS software version 6.30 and the {\it Suzaku} calibration database (CALDB) files released in 2016. We used the reprocessing tool \texttt{aepipeline} to reprocess the observations, estimated the non-X-Ray background (NXB) with \texttt{xisnxbgen} \citep{tawa08}, and created the rmf and arf response files with \texttt{xisrmfgen} and \texttt{xisarfgen}. 

\subsection{Spectral Fitting}
\label{sec:specanal}
We used XSPEC version 12.13.0e \citep{arnaud96} and AtomDB v3.0.10 \footnote{This version of AtomDB is currently unreleased. In it, the atomic transition values for K-$\alpha$ and -$\beta$ lines in Fe$+$15 and below were made more accurate with lab data. This change is significant for plasmas with kT$_e\gtrsim$ 2~keV and $\tau_{e} \lesssim 5 \times 10^{10}$ s cm$^{-3}$.} \citep{smith01,foster12} to fit the 0.6--8.0~keV X-Ray spectrum of Kepler's SNR. This bandpass includes strong K-$\alpha$ lines from shock-heated ejecta species O, Ne, Mg, Si, S, Ar, Ca, Cr, Mn, Fe, and Ni, as well as many Fe L-$\alpha$ lines around 1~keV and prominent Fe K$\beta$ lines at $\sim$7.1~keV. Although we only report final ejecta mass ratios of elements Si and more massive, we included down to 0.6~keV in order to constrain the column density towards Kepler's SNR, model out the shocked swept-up surrounding material, and include the Fe-L emission lines.

Our model includes a component that accounts for interstellar absorption (\texttt{tbabs}; \citealt{wilms00}), a dedicated swept-up ISM/CSM component (\texttt{vpshock}), a non-thermal component (\texttt{srcut}; $\alpha$=-0.71 and roll-off frequency $\approx$1--$3 \times 10^{17}$~Hz; \citealt{delaney02,green19,nagayoshi21}) and multiple non-equilibrium ionization (nei) thermal plasmas to capture ejecta emission. Each component is multiplied by a gaussian smoothing model \texttt{gsmooth} to account for Doppler-broadened emission.

As Type Ia SN ejecta are dominated by heavier elements, we followed the technique of past studies (e.g., \citealt{katsuda15}) that use plasma components dominated by elements heavier than hydrogen. We included at least one IME-dominated and one Fe-dominated ejecta component characterized by abundances [Si/H] and [Fe/H] fixed at 10$^5$ times solar, respectively. Our best fits were achieved with two IME-dominated components represented by the plane-parallel shock models \texttt{vpshock} and one Fe-dominated component represented by a singly-shocked plasma model \texttt{vvnei}---appropriate for a more homogeneous, smaller-in-scale plasma. Using more than these three ejecta-dominated plasma components, or using different combinations of plasma model types, did not improve our fit.

Our final model was \texttt{tbabs}*(\texttt{gsmooth}*\texttt{vpshock}$_{\rm CSM}$ + \texttt{gsmooth}*\texttt{vpshock}$_{\rm ej1}$ + \texttt{gsmooth}*\texttt{vpshock}$_{\rm ej2}$ + \\ \texttt{gsmooth}*\texttt{vvnei}$_{\rm ej3}$ + \texttt{srcut}). We left the column density N$_{\rm H}$ as a free parameter with initial value of $6.4 \times 10^{21}$ cm$^{-2}$ \citep{katsuda15}. We allowed the electron temperature, normalization, ionization timescale, redshift (representing bulk ejecta velocity), and gaussian smoothing in each plasma component, as well as the roll-off frequency and normalization of the non-thermal component, to vary. We note that the centroid of the Fe K-$\beta$ feature at $\sim$7.1~keV and its flux ratio to the Fe K-$\alpha$ feature placed a strict constraint on the ionization timescale of the Fe-dominated plasma: n$_e$t $\approx3.7\times 10^{9}$ cm$^{-3}$ s.

In the shocked CSM/ISM component, we set the abundance of N to 3.3 \citep{blair91, katsuda15} and allowed the abundances of O, Ne, Mg, Si, S, Ar, Ca, and Fe to vary between 0.1 and 1 times solar. In the two IME-dominated components, we allowed the abundances of elements heavier than nitrogen to vary, tied Ni to Fe, and froze [Si] to 10$^5$ as described previously. Additionally, we tied all abundances---except for Fe---of the two IME-dominated components (ej1 and ej2) together. In the Fe-dominated component (ej3), we allowed the abundances of Cr, Mn, and Ni to vary while freezing [Fe] to 10$^5$. The abundances of all unmentioned elements are set to solar.

An example of our spectral fit is shown in Figure~\ref{fig:specfit}. As shown, it contains many strong residuals when compared to the data, resulting in high reduced-$\chi^2$ values of 2.5--3.5. However, our fits still capture and can constrain features of the spectra most important for our analysis. The broad spectral shapes from thermal and non-thermal continuum emission---shapes which constrain component normalization and temperature---are well fit, as are many of the strong K-$\alpha$ lines. Much of the poor statistical fit comes from low energy, high signal-to-noise bandpasses that don't contain emission lines from heavy elements. For example, over the 4.5--8.0~keV bandpass (containing IGE K-$\alpha$ lines), the reduced-$\chi^2$ of our fits decrease to $\sim$1--1.3. Additionally, we note that extracted {\it Suzaku} spectra do not include systematic uncertainties associated with telescope calibration; thus the true photon uncertainties are larger than shown.

\subsubsection{Accounting for X-Ray Calibration Uncertainties}
\label{subsubsec:cal_uncerts}

\begin{figure*}
\begin{center}
\textbf{Effective Area Curves} \\
\includegraphics[width=0.48\textwidth]{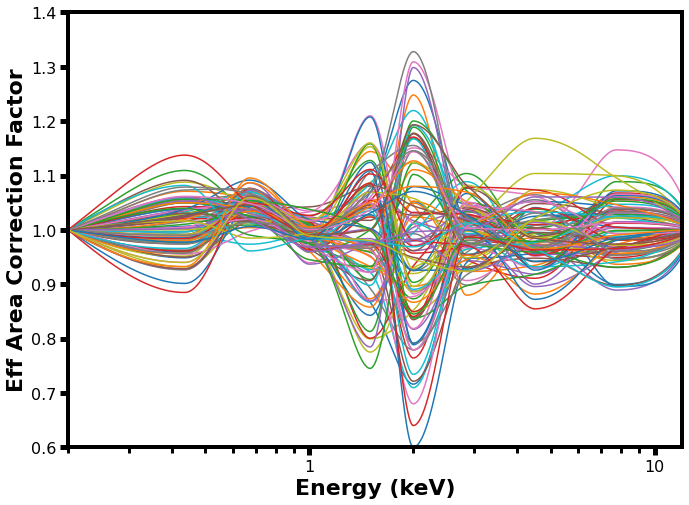}
\includegraphics[width=0.495\textwidth]{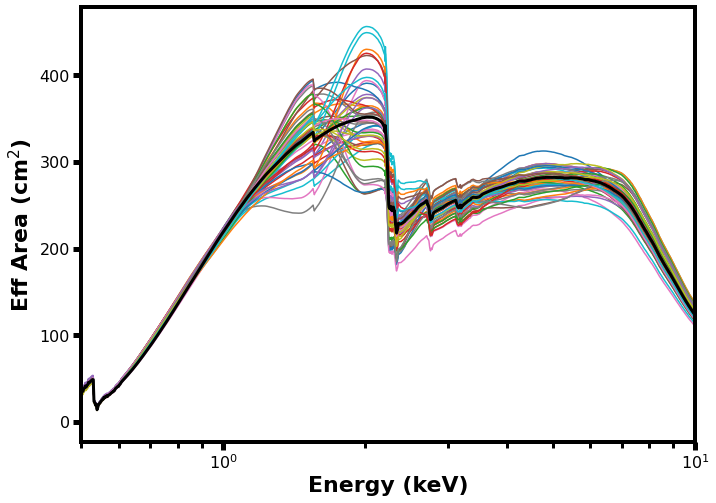}
\end{center}
\vspace{-5mm}
\caption{\footnotesize{(Left): {\it Suzaku} effective area correction curves generated from uncertainty priors reported in \cite{marshall21} and our 
Table~\ref{table:effareauncerts}. 
(Right): the resulting effective area curves, generated via multiplying the correction curves by the appropriate {\it Suzaku} ancillary response file. The right plot only shows 50 mock effective area curves to reduce clutter.}}
\label{fig:EffAreaCurves}
\end{figure*}

To address the issue of telescope uncertainties, we follow a procedure similar to that of \cite{lee11} and \cite{xu14} who accounted for the effects of systematic calibration uncertainties in spectral fits to {\it Chandra} data. Using bandpass-specific {\it Suzaku} effective area calibration uncertainties reported by \cite{marshall21} and presented in Table~\ref{table:effareauncerts}, we generated 100 effective area correction curves. At the center of each bandpass, we generated a number drawn from a normal distribution with variance given by reported the 1-$\sigma$ uncertainty. We then smoothed this curve via the Python {\it scipy} \texttt{PchipInterpolator} algorithm---an interpolator designed to prevent overshooting the data---and multiplied these curves by the base {\it Suzaku} effective area curve to generate 100 mock effective area curves. Examples of these mock effective area curves are shown in Figure~\ref{fig:EffAreaCurves}. Finally, we used these 100 mock curves to fit 100 spectra, and took the resulting spread as reflective of the telescope calibration uncertainties.

\subsection{MCMC based fitting}
\label{subsec:MCMC}

\begin{figure*}
\begin{center}
\textbf{MCMC plots} \\
\includegraphics[width=\textwidth]{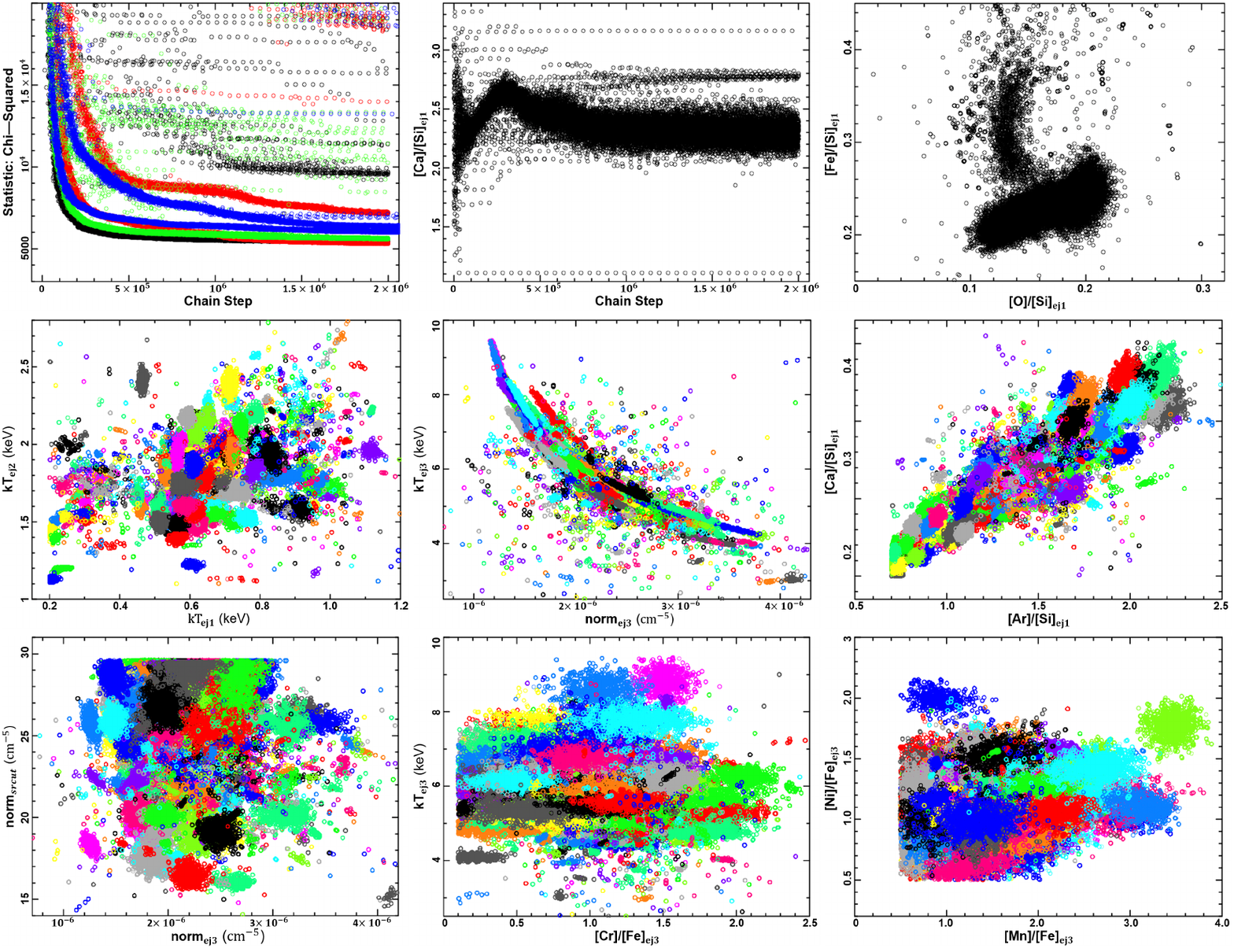}
\end{center}
\vspace{-5mm}
\caption{\footnotesize{The top row shows: (left) the evolution of the total fit statistic for a few different effective area curves, (middle) the evolution of the best-fit [Ca]/[Si] abundance ratio for a single MCMC run, and (right) the 500 final best-fit [Fe]/[Si] vs [O]/[Si] abundance ratios from an ejecta plasma component for a single MCMC run. The middle row shows sample final parameter values for our 100 MCMC runs; each color reflects the results from a single MCMC run. The bottom row shows parameters that are dominated by higher-energy, lower signal-to-noise emission. As such, these parameters can become unconstrained or obviously improperly fit, as shown by the sharp cutoffs and large spreads. See the text in Section~\ref{subsubsec:Fe-groups} for discussion on how we dealt with this issue.}}
\label{fig:MCMC_plots}
\end{figure*}

To fully investigate the parameter space, we used the XSPEC \texttt{chain} command to perform Markov Chain Monte Carlo based fitting. Given the large ($\sim$35) number of free parameters in our model, we used the Goodman-Weare algorithm with 500 walkers (i.e., different parameter combinations) to ensure sufficient spread. We used uniform distributions for initial walkers, with ranges determined by initial rough fits and physical plausibility. We found that $\gtrsim$5000 steps per walker were needed to obtain a Geweke convergence statistic of $<$0.05. See the top row of Figure~\ref{fig:MCMC_plots} for examples of fit statistic evolution and final MCMC parameter values.

For a given model parameter, each MCMC run produced a spread of best-fit values that we condensed into a mean value P and 1-$\sigma$ statistical uncertainty $\sigma_{i, {\rm stat}}$. We averaged this uncertainty over the 100 runs to obtain a final statistical uncertainty $\overline{\sigma_{\rm stat}}$. The {\it Suzaku} calibration uncertainty ($\sigma_{\rm cal}$) was reflected in the variance of the 100 average parameter values. We then combined these errors in quadrature to get the total best-fit parameter uncertainty:
\begin{equation}\sigma_{\rm param}^2 = \overline{\sigma_{\rm stat}}^2 + \sigma_{\rm cal}^2\end{equation}
The middle and bottom rows in Figure~\ref{fig:MCMC_plots} show a selection of final parameter values, where each color represents the spread from an individual MCMC runs.

\begin{deluxetable}{ccccccc}[!t]
\tablecolumns{7}
\tablewidth{0pt} 
\tablecaption{{\it Suzaku} XIS Effective Area Uncertainty Priors (\%)}  
\tablehead{\multicolumn{7}{c}{Energy Bands (keV)} \\ \colhead{0.5--0.8} &\colhead{0.8--1.2} &\colhead{1.2--1.8}  &\colhead{1.8--2.2}  &\colhead{2.2--3.5}  &\colhead{3.5--5.5}  &\colhead{5.5--10}}
\startdata
15\tablenotemark{a} & 10\tablenotemark{a} & 10 & 15 & 5 & 5 & 5 \\ 
\enddata
\tablenotetext{a}{\cite{marshall21} investigated further and obtained posterior values for effective area correction factors: 1.05$\pm$0.03 for 0.54--0.8~keV and 0.99$\pm$0.02 for 0.8--1.2~keV. We use these values.}
\label{table:effareauncerts} 
\end{deluxetable}

\subsubsection{Cr, Mn, and Ni abundances}
\label{subsubsec:Fe-groups}

We found that the abundances of Cr, Mn, and Ni were often unconstrained in MCMC fits to the entire 0.6--8.0~keV spectra, likely due to the vastly larger signal-to-noise values present in the lower energies of the spectra that dominate $\chi^2$ minimization. The effect can be seen in the bottom row of Figure~\ref{fig:MCMC_plots}, where the normalization of the synchrotron component and Fe-group ejecta abundances often reached the limits set by us and generally exhibit a large spread within each MCMC run. Plotting these limit-reached fits revealed that the synchrotron component's contribution was overestimated, resulting in model flux that clearly exceeded the observed spectrum---namely the Cr/Mn/Ni K-$\alpha$ emission.

As such, we perform additional MCMC fitting on a restricted 4.5--8.0~keV bandpass. This bandpass was chosen to contain little emission from swept-up material and the IME-dominated plasmas, but still maximize the continuum present in order to properly account for its contribution. Fits to this region produced reduced-$\chi^2$ values of $\sim$1--1.3. We froze most parameters values to those in the full-bandpass fits, but allowed the Fe-dominated plasma component normalization and IGE abundances, along with the normalization of the synchrotron component, to vary. We then use the reported Cr, Mn, and Ni spreads as the ``true'' best-fit parameter values and uncertainties. We note that these results matched well with the spreads of the full-bandpass MCMC runs if we exclude fits that clearly overestimated the high-energy flux. Our final best-fit spectral parameters and their uncertainties are presented in Appendix~\ref{appen:MRs}.

\subsection{Calculating Ejecta Mass Ratios}
\label{subsec:MRCalcs}
In this section, we discuss a significant limitation of using multi-component plasma models: adding ejecta masses from different plasma components requires assumptions about the unknown emitting volumes of each component. We can describe this as an unknown filling factor $f$ multiplied by the total volume: V$_{\rm emit} = \rm{V}_{ \rm tot} \times f$. The total mass of an element X (full derivation presented in Appendix~\ref{appen:MRs}) is given by
\begin{equation}
    M_{\rm X, tot} = M_{\rm X, ej1}f_{\rm ej1}^{0.5} + M_{\rm X, ej2}f_{\rm ej2}^{0.5} + M_{\rm X, ej3}f_{\rm ej3}^{0.5}
\end{equation} 
where M$_{\rm X,ejY}$ is the mass contribution from plasma component Y assuming a filling factor of 1, i.e., uniformly filling the total volume of Kepler's SNR. We can make reasonable assumptions about how the filling factors of each plasma relate to each other and thus simplify the above equation. 
\begin{equation}
    M_{\rm X, tot} = \big(M_{\rm X, ej1} + \alpha M_{\rm X, ej2} + \beta M_{\rm X, ej3}\big)f_{\rm ej1}^{0.5}
\end{equation} 
where $\alpha=(f_{\rm ej2}/f_{\rm ej1})^{0.5}$ and $\beta=(f_{\rm ej3}/f_{\rm ej1})^{0.5}$. This allows us to obtain final ejecta mass ratio estimates as the $f_{\rm ej1}^{0.5}$ terms will cancel out during division.

We investigated four methods of estimating the different filling factors:
\begin{enumerate}
\item Equal Emitting Volumes
\item Electron Pressure Equilibrium: P${_e}$=$n_ekT_e$ is constant between ejecta components, leading to:
\begin{equation} f_m^{0.5} = \frac{(\eta_{e}T_{e})_m}{(\eta_{e}T_{e})_n} f_n^{0.5} \end{equation}
where $\eta_e$ is the electron number density $n_e$ calculated assuming a filling factor of 1.
\item Linking plasmas to specific regions of the SNR. We use the results of \cite{katsuda15}, who linked IME-dominated emission to a shell region 0.85--0.97 times Kepler's forward shock radius R$_{\rm FS}$ and Fe-dominated emission to a shell region from the reverse shock ($\sim$0.7R$_{\rm FS}$) to 0.85R$_{\rm FS}$. The two IME-dominated plasmas are assumed to be in electron pressure equilibrium fill the entire outer shell, and the Fe-dominated plasma fills the inner shell: 
\[f_{\rm ej1}+f_{\rm ej2}=(0.97^3-0.85^3) \rm{\ \ ; \ \ } P_{_{\rm ej1}} = P_{_{\rm ej2}} \]
\begin{equation} f_{\rm ej3}=(0.85^3-0.7^3)
\end{equation}
\item As above, but enforced pressure equilibrium between the two shells. \begin{equation}P_{\rm ej3}=P_{\rm ej1}+P_{\rm ej2}=2P_{\rm ej1}\end{equation}
\end{enumerate}

\begin{figure}
\begin{center}
\includegraphics[width=\columnwidth]{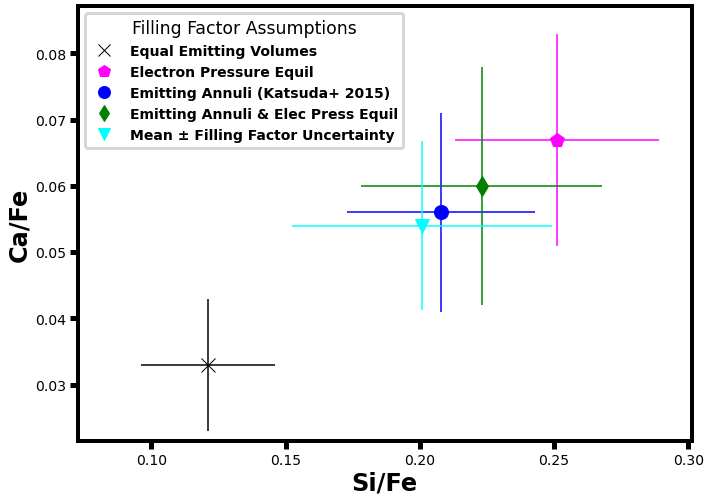} 
\includegraphics[width=\columnwidth]{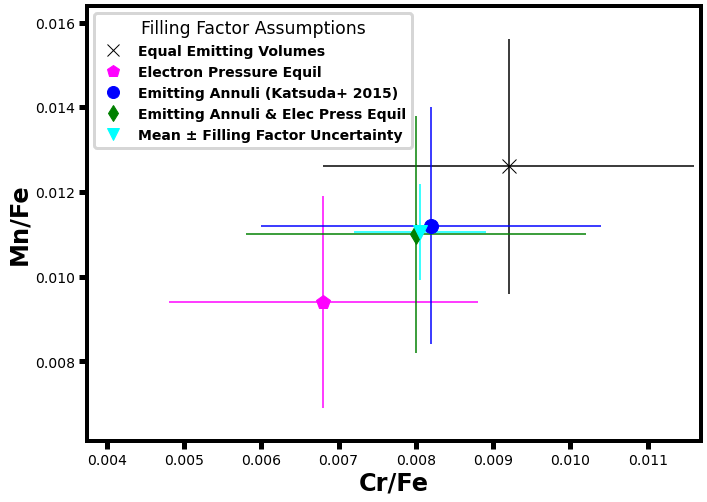} 
\end{center}
\vspace{-5mm}
\caption{\footnotesize{Each non-cyan data point is a mass ratio calculated using a different filling factor assumption, and the cyan triangle is the average of these four estimates. The error bars on the non-cyan points are propagated from best-fit parameter uncertainties. The error bars on the cyan data point are the standard deviation of those four estimates---reflecting filling factor uncertainty. }}
\label{fig:diff_methods}
\end{figure}

The full derivations of these equations are presented in Appendix~\ref{appen:FillFs}

With current measurements, it is not possible to know which, if any, filling factor assumption is correct, so we estimate mass ratios using each of the four methods and quantify the spread of these estimates. The effects of different filling factor assumptions on calculated ejecta mass ratios are shown in Figure~\ref{fig:diff_methods}. As shown, the choice of filling factor can change the final mass ratio estimate, and thus accounting for the various possible filling factors introduces an additional uncertainty.

\subsubsection{Multiple Imputation of Filling Factors}
\label{subsubsec:Mul_Imp}
To obtain final mass ratios estimates and uncertainties, we combine our MCMC-derived results with the effects of assuming different filling factors. As four different assumptions is too few to simply add the errors in quadrature, we use the Multiple Imputation technique---a technique designed to handle missing data \citep{rubin87,schafer97}---to combine the errors. The total uncertainty on our mass ratio estimates is thus given by 
\begin{equation}\sigma_{\rm tot}^2 = \overline{\sigma_{\rm fit}^2}+\bigg(1+\frac{1}{M} \bigg)\sigma_{\rm vol}^2\end{equation}
\\
where M is the number of imputations (4 in our case), $\overline{\sigma_{\rm fit}}$ is the average fit uncertainty from the four mass ratio estimates, and $\sigma_{\rm vol}$ is the standard deviation of the mass ratio estimates calculated using our 4 filling factor assumptions.

To account for the small value of M, we use a t-distribution to find the appropriate 68.3\% confidence interval with the degrees of freedom given by
\begin{equation}{\rm degrees\ of\ freedom} = (M-1)\bigg(1+ \frac{M \sigma_{\rm fit}^2}{(M+1)\sigma_{\rm vol}^2}\bigg)^2\end{equation}

\section{Results}
\label{sec:results}
Our final mass ratios and their 68.3\% confidence intervals are reported in Table~\ref{table:MRs}. The uncertainties of our estimates strongly depend on the relative contribution of the ejecta from each component:
\begin{enumerate}
\item Mass ratios involving elements whose abundances were tied between components (Si/S, Ar/S, and Ca/S) have 10--20\% errors that solely originate from propagated fit errors. The filling factor terms cancel out for these elements.
\item Mass ratios involving elements whose masses were mostly from different plasma components (i.e., IMEs/Fe) have $\sim$40\% errors, with a larger contribution from the filling factor uncertainty.
\item Mass ratios whose elements were predominantly from the same component (Cr/Fe, Mn/Fe) have $\sim$30\% errors, with the filling factor uncertainty contributing a small but non-zero amount. For Ni/Fe, as its abundance was tied to the Fe abundance for the two IME-dominated components, it is reasonable to assume that its full uncertainty should be closer to the other IGE/Fe mass ratios.
\end{enumerate}

\begin{deluxetable}{lccrc}[!t]
\tablecolumns{4}
\tablewidth{0pt} 
\tablecaption{Calculated Mass Ratios in Kepler's SNR\label{table:MRs}} 
\tablehead{ \colhead{Element } & \colhead{Final} & \colhead{Summed}&\colhead{$\sigma_{\rm vol}$\tablenotemark{b}} &\colhead{$\sigma_{\rm fit}$\tablenotemark{c}} \\
\colhead{Ratio} &\colhead{Value} & \colhead{Sq Err\tablenotemark{a}} & & }
\startdata
Si/Fe & {\bf 0.199 $\pm$ 0.082} & 0.068 & 0.055   & 0.040    \\
Si/S  &  {\bf 1.25 $\pm$ 0.117} & 0.117 & 0     & 0.117 \\
S/Fe  & {\bf 0.160 $\pm$ 0.067} & 0.056 & 0.044   & 0.036 \\
Ar/Fe & {\bf 0.048 $\pm$ 0.022} & 0.020 & 0.013   & 0.014   \\
Ar/S  & {\bf 0.299 $\pm$ 0.073} & 0.073 & 0     & 0.073 \\
Ca/Fe & {\bf 0.053 $\pm$ 0.023} & 0.021 & 0.014   & 0.016   \\
Ca/S  & {\bf 0.333 $\pm$ 0.080} & 0.080 & 0   & 0.080 \\
Cr/Fe & {\bf 0.0082 $\pm$ 0.0024} & 0.0023 & 0.0010 & 0.0021\\
Mn/Fe & {\bf 0.0114 $\pm$ 0.0031} & 0.0030 & 0.0014 & 0.0026 \\
Ni/Fe & {\bf 0.0550 $\pm$ 0.0093\tablenotemark{d}} & 0.0093 & 0.0014 & 0.0092  \\ \hline
\enddata
\tablenotetext{a}{Not accounting for Multiple Imputation}
\tablenotetext{b}{Uncertainty from unknown plasma component filling factors}
\tablenotetext{c}{Uncertainty from spectral fitting: combined statistical (photon) \& systematic (telescope calibration) uncertainties.}
\tablenotetext{d}{The Ni abundance was tied to Fe for the two IME-dominated components (ej1 and ej2).
}
\vspace{-8mm}
\end{deluxetable}

\subsection{Unshocked Ejecta in Kepler's SNR}
\label{subsec:unshocked_obs}
Table~\ref{table:MRs} presents the total mass ratios estimated using all plasma components; i.e., calculated from all shocked X-Ray emitting material. However, the inner region of Kepler's SNR (r$\lesssim0.7$R$_{\rm Kep}$; \citealt{katsuda15}) is unshocked and is not emitting in X-Rays; thus ejecta in these interior regions are not captured by our analysis.

The interior of Type Ia SNRs should contain ejecta dominated by IGEs due to stratification of ejecta \citep{badenes06, katsuda15}. As such, our estimated IME/Fe ratios are likely overestimates. However, the relation between IGE/Fe mass ratios and percentage of unshocked ejecta is less straightforward. Some models (mainly single detonations; e.g., \citealt{maeda10}) suggest that Mn is formed in the innermost, densest regions of the explosion, while other models (mainly double-detonations, e.g., \citealt{lach20}) suggest that Mn is formed in the He-shell detonation and thus will be in the outer layers of expanding ejecta.

\cite{katsuda15} estimated that the outer $\sim$87\% of the ejecta in Kepler's SNR is shocked, while a more conservative estimate based purely on the shocked vs. unshocked volumes is that the outer $(0.97^3-0.7^3)/0.97^3\approx$60\% of ejecta is shocked. We discuss the effects of incomplete reverse-shock propagation on observed ejecta mass ratios in Section~\ref{sec:simuls} when we compare our results to various Ia simulations.

\begin{deluxetable*}{lccccccc}[!h]
\tablecolumns{4}
\tablewidth{0pt} 
\tablecaption{Mass Ratio Comparisons \label{table:prevMRs}} 
\tablehead{ \colhead{Element } & \colhead{Our} & \colhead{Katsuda15}  &\colhead{MR17}  &\colhead{Sato20} &\colhead{Park13} & {Yamaguchi15}\\
\colhead{Ratio} &\colhead{Values}
&\colhead{Values\tablenotemark{a,b}} & \colhead{Values\tablenotemark{b}} &\colhead{Values\tablenotemark{b,d}} &\colhead{Values\tablenotemark{e, f}}
&\colhead{Values\tablenotemark{e}}}
\startdata
Si/Fe & 0.199 $\pm$ 0.082 
& 0.056 $\pm$ 0.059 &  & \\
Si/S &  1.25 $\pm$ 0.12 
& 0.851 $\pm$ 1.243 & &  \\
S/Fe & 0.160 $\pm$ 0.067 
& 0.066 $\pm$ 0.073 & &  \\
Ar/Fe &0.048 $\pm$ 0.022 
& 0.017 $\pm$ 0.026 & &  \\
Ar/S & 0.299 $\pm$ 0.073 
& 0.255 $\pm$ 0.470  & 0.279$_{-0.017}^{+0.010}$&  \\
Ca/Fe &  0.053 $\pm$ 0.023 
& 0.030 $\pm$ 0.036  & & \\
Ca/S & 0.333 $\pm$ 0.080 
& 0.447 $\pm$ 0.713 & 0.283$_{-0.023}^{+0.016}$&  \\
Cr/Fe & 0.0082$\pm$0.0024 
& & 0.008$_{-0.005}^{+0.007}$&\\
Mn/Fe & 0.0114$\pm$0.0031 
& & & & & 0.01$^{+0.006}_{-0.0035}$  \\
Mn/Cr & 1.38 $\pm$ 0.47 
& & & & 0.77$^{+0.53}_{-0.31}$ \\
Ni/Fe & 0.0550$\pm$0.0093 
& 0.155\tablenotemark{c} & & & 0.06$^{+0.04}_{-0.02}$ & 0.045$^{+0.03}_{-0.02}$\\ \hline
\cutinhead{Mass Ratios From ej3 Only} \hline
Ca/Fe &  
&   & & 0.025 $\pm$ 0.008\\
Cr/Fe & 0.0108$\pm$0.0025 
&  &  & 0.027 $\pm$ 0.01\\
Mn/Fe & 0.0150$\pm$0.0031 
&  & & 0.019$_{-0.005}^{+0.04}$\\
Ni/Fe & 0.0586$\pm$0.0083 
& & & 0.15 $\pm$ 0.07 & &  \\ 
\enddata
\tablenotetext{a}{Shocked ejecta mass ratios. If we account for their estimate of unshocked, IGE-dominated ejecta, all IME/Fe ratios would decrease by $\sim$13\%.}
\tablenotetext{b}{Mass ratios were calculated using the old AtomDB Fe-K ionization emissivities, which overestimates IME/Fe mass ratios by $\sim$15\% and IGE/Fe mass ratios by $\sim$30\%.}
\tablenotetext{c}{Ni abundance was tied to Fe abundance} 
\tablenotetext{d}{Fit a Fe-rich region only; averaging over the 3 best-fit electron temperatures.}
\tablenotetext{e}{Calculated using flux and emissivity ratios instead of best-fit abundances.}
\tablenotetext{f}{Emissivities calculated ``using the new atomic data (K. A. Eriksen et al., in preparation) generated for this project with Flexible Atomic Code (Gu 2008).''}
\end{deluxetable*}

\section{Previous Observational Studies}
\label{sec:pastwork}

\label{subsec:prevpapers}
The main observational studies we compare our work to are \cite{katsuda15}, \cite{mr17}, \cite{sato20}, and \cite{park13} (expanded on by \citealt{yamaguchi15}), each of whom analyzed Kepler's SNR slightly differently. We report mass ratios calculated from their works in Table~\ref{table:prevMRs}.

\subsection{Global Analysis of IMEs and Fe}

The paper by \cite{katsuda15} is the most similar to ours: a full X-Ray spectral analysis of the entire SNR using essentially the same spectral model. However, their spectral fits do not include Cr or Mn abundances, and their Ni and Fe abundances are tied for all spectral components, whereas we thaw Mn, Cr, and Ni in our Fe-dominated component. They estimated the filling factors of their plasma components using the radial-shell method described in Section~\ref{subsec:MRCalcs} (method \#3) and report uncertainties that are much greater than the statistical uncertainties.

Our calculated mass ratios are mostly consistent with those reported in \cite{katsuda15}. Our IME/Fe estimates are systematically higher than theirs by factors of 2--4 (0.5--1.5-$\sigma$ differences), while our IME/S mass ratio estimates have $<$1-$\sigma$ differences. We note that the updated atomic data we used has smaller Fe emissivities and thus {\it should} result in smaller X/Fe mass ratios. However, all of our calculated IME/Fe ratios are higher than theirs: a puzzling detail.

\cite{katsuda15} concluded that Kepler's SNR was an overluminous event interacting with massive CSM from the progneitor system based on their relatively high IGE-to-IME mass ratio. Our findings partially support that conclusion but are also consistent a normally energetic explosion.

\subsection{Global Separate Analysis of IMEs and IGEs}
\cite{mr17} also analyzed the global spectrum of Kepler's SNR with {\it Suzaku} data, and they included Cr, Mn, and Ni as free parameters. However, they analyzed the 2--5~keV (Si, S, Ar, and Ca K$\alpha$ emission lines) and 5--8~keV (Cr, Mn, Fe, and Ni lines) bandpasses separately, fitting each with only a single \texttt{vvpshock} plus a continuum component. This method results in a loss of constraining information from different energies. For example, when we fit Kepler's 5--8~keV bandpass with a single thermal plus power-law model---without ensuring consistency with a broader 0.6--8.0~keV fit---our fit drastically overestimated flux from lower energies.

\cite{mr17} reported Ar/S, Ca/S, and Cr/Fe mass ratios consistent with ours to within $\lesssim$1-$\sigma$. Notably, our Cr/Fe mass ratio estimate has a a factor of $\sim$2.5 smaller uncertainty than theirs; we suggest that using the Fe-L lines near $\sim$1~keV aided in constraining the plasma properties of the Fe-dominated component. 

Compared to the 1D explosion models of \cite{yamaguchi15}, the results of \cite{mr17} favored higher detonation densities and 0.5--2Z$_\odot$ progenitors. Additionally, they found that many multi-dimensional Ia models systematically underpredicted the Ca/S ratios observed in Ia SNRs and suggested that a $\sim$90\% attenuation of the $^{12}$C+$^{16}$O rate was appropriate to bring Type Ia simulations in line with matching observations. Our findings support this conclusion.

\subsection{Analysis of IGEs in a Fe-rich structure}
\cite{sato20} estimated IGE mass ratios within a Fe-rich substructure in the southwest of Kepler's SNR composing $\sim$3\% of its volume. They analyzed {\it Chandra} data of that region, fitting a thermal plus power-law model to the 3.5--9~keV spectrum. Similar to the study by \cite{mr17}, their method of fitting a single thermal plasma component to a restricted bandpass might have resulted in a model that does not properly represent the plasma(s) present. However, given their restriction to a small---and thus more homogeneous---region, their single thermal plasma model is more likely to accurately represent the true plasma properties.

We compare their results to mass ratios calculated using our Fe-dominated ejecta component. Our estimated Mn/Fe mass ratio is within 1-$\sigma$ of theirs, but our Cr/Fe and Ni/Fe mass ratio estimates differ from theirs by 1.5-$\sigma$. However, if we correct for the updated AtomDB emissivities, the reported IGE/Fe mass ratios of \cite{sato20} would decrease by $\sim$30\% and be within 1-$\sigma$ of our estimates. We note that our reported uncertainties are significantly smaller than theirs, even though we included multiple additional sources of uncertainty. We suggest that this is a result of our use of a wider bandpass, enabling higher precision in the best-fit properties of the Fe-dominated plasma.

\cite{sato20} concluded that their mass ratio estimates were consistent with being synthesized via incomplete Si-burning of a 
Z$>$1.3Z$_\odot$ 
metallicity progenitor, but could not distinguish between near- and sub-M$_{\rm Ch}$ progenitors. Our results agree with their findings.

\subsection{Global Flux Study of IGEs}
\cite{park13} measured the flux of IGE emission lines and used element emissivities to convert those fluxes into masses or mass ratios. Similar to the paper by \cite{mr17}, they analyzed the 5--8~keV bandpass of the entirety of Kepler's SNR using {\it Suzaku} observations, fitting the spectrum with a powerlaw plus 5 gaussians to measure K-$\alpha$ emission from Fe-group elements. 

\cite{park13} reported a Mn/Fe mass ratio of 0.77$^{+0.53}_{-0.31}$, suggesting a super-solar progenitor metallicity of 
Z$\approx3_{-2}^{+5}$Z$_\odot$ 
according to the 
power-law relation M$_{\rm Mn}$/M$_{\rm Cr}$ = 5.3Z$^{0.65}$ 
of \cite{badenes08} and thus a relatively prompt channel for a near-M$_{\rm Ch}$ explosion. They also reported a Ni/Fe mass ratio of 0.06$^{+0.04}_{-0.02}$, in agreement with near-M$_{\rm Ch}$ delayed detonation models \citep{iwamoto99}.

We note that our while our estimated Ni/Fe mass ratio  matches theirs, our Mn/Cr ratio is almost twice as large as theirs, although this is only a $\sim$1-$\sigma$ difference. According to the Mn/Cr-metallicity power relation of \cite{badenes08}, our Mn/Cr ratio of $\sim$1.38 suggests an extreme progenitor metallicity of Z$\approx$0.12$_{-0.06}^{+0.07}$ (Z/Z$_\odot\approx7_{-3.5}^{+5}$). Given that this metallicity is extremely high, our findings suggest that neutronization due to carbon simmering may be an important factor for Kepler's SNR.

\subsubsection{Follow-up Flux Studies}
\cite{yamaguchi15} took the measured line fluxes from \cite{park13} and used updated atomic data to estimate a Mn/Fe mass ratio of 0.01$^{+0.006}_{-0.0035}$ and a Ni/Fe mass ratio of 0.055$^{+0.03}_{-0.02}$. They compared these values to various 1D near- and sub-M$_{\rm Ch}$ explosion models and found that either near-Ch DDT models with Z$\approx$1.8Z$_\odot$ and medium-low detonation densities ($\sim$1.3e7 g cm$^{-3}$) or sub-Ch models with Z$\lesssim$Z$_\odot$ are favored. 

We note that, although our estimates match the results of \cite{yamaguchi15}, we found that many Type Ia models that matched our estimates of Mn/Fe and Ni/Fe didn't match well with our estimate for the Cr/Fe mass ratio (see Section~\ref{sec:simuls}). As \cite{yamaguchi15} didn't report a Cr/Fe mass ratio, this limits the significance of their proposed associations.

Last of all, studies such as \cite{mr16,piersanti22} suggested that pre-explosive accretion and carbon simmering can significantly affect nucleosynthesis. \cite{piersanti22} derived an equation to relate neutron excess to progenitor metallicity in near-$M_{\rm Ch}$ Type Ia SNe. Using the IGE flux data of Kepler's SNR analyzed in \cite{park13,yamaguchi15} and the derived Mn/Cr mass ratio, they concluded that Z$_{\rm Kepler}$=0.0373=2.71Z$_\odot$. Our estimated Mn/Cr mass ratio is consistent with the results of those studies, suggesting a high progenitor metallicity.

\section{Ia Simulations and Origin Scenarios}
\label{sec:simuls}
In this section, we discuss the nucleosynthesis yields of various near- and sub-M$_{\rm Ch}$ Type Ia simulations and compare them to our estimates, aiming to place constraints on the origin of Kepler's SNR and indicate parameter spaces for future simulations to investigate. 

\begin{figure*}
\begin{center}
\textbf{Near-M$_{\rm Ch}$ Type Ia Model Nucleosynthesis Results} \\
\includegraphics[width=0.95\textwidth]{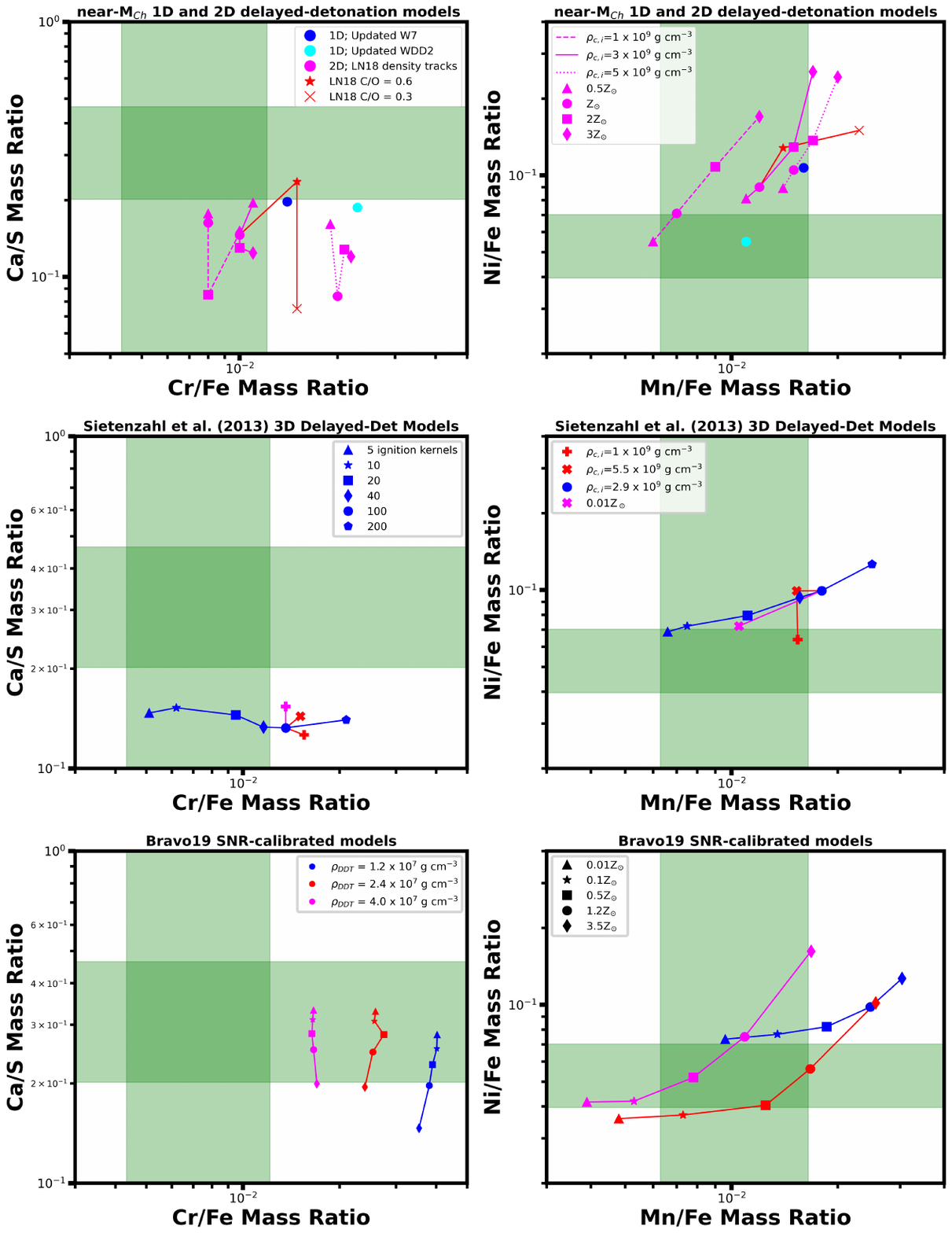}
\end{center}
\vspace{-5mm}
\caption{\footnotesize{Our estimated mass ratios for ejecta in Kepler's SNR (green bars, representing 90\% confidence intervals) compared to the results of various near-M$_{\rm Ch}$ Delayed Detonation nucleosynthesis models. LN18 refers to \cite{leung18}; Maeda10 refers to \cite{maeda10}; and Bravo19 refers to \cite{bravo19}}. Each row of plots is a pair, with labels split between the two plots.}
\label{fig:NearCh_simulations}
\end{figure*}

\subsection{Near-M$_{\rm Ch}$ Explosion Models}
In Figure~\ref{fig:NearCh_simulations}, we present the 90\% confidence intervals of our mass ratio estimates compared to the nucleosynthesis yields of various near-M$_{\rm Ch}$ simulations. We plot the Ca/S and all 3 IGE/Fe mass ratios, but not the IME/Fe mass ratios as our estimates matched all simulations to roughly the same degree.

The top row of Figure~\ref{fig:NearCh_simulations} shows results from various 1D and 2D models. The blue and cyan points are from the classic 1D near-M$_{\rm Ch}$ W7 and WDD2 models, representing pure deflagration and a deflagration-to-detonation transition respectively \citep{nomoto84,iwamoto99} and using updated nucleosynthesis data of \cite{nomoto18,mori18}. The magenta and red data points are from 2D near-M$_{\rm Ch}$ deflagration-to-detonation simulations of \cite{leung18}. They investigated the effects of initial central density (a proxy for WD mass: 1.30--1.38M$_\odot$), metallicity (Z=0--5Z$_\odot$), and progenitor C/O mass ratios (0.3--1.0). 
As shown, nearly all of these models underpredict Ca/S and overpredict Cr/Fe and Ni/Fe. The models that come the closest are the low-density, sub-solar metallicity, and high C/O models of \cite{leung18}.

The middle row of Figure~\ref{fig:NearCh_simulations} shows the nucleosynthesis yields for 3D near-M$_{\rm Ch}$ DDT models of \cite{seitenzahl13}. They investigated the effects of metallicity, central density, and distribution of ignition kernels. Compared to our estimates, all of their models underproduce Ca/S and overproduce Ni/Fe. However, a model with a small number of ignition kernels (10--40, corresponding to the production of $\sim$1--0.7M$_\odot$ of $^{56}$Ni, respectively), lower metallicity, and/or lower WD central density might match our IGE mass ratio estimates. This specific parameter space was not reported in their paper.

The bottom row of Figure~\ref{fig:NearCh_simulations} shows the nucleosynthesis results of \cite{bravo19}. Based on the conclusion of \cite{mr17} that a reduced $^{12}$C$+^{16}$O reaction rate was necessary to match observed SNR Ca/S mass ratios, they computed a new set of 1D SN Ia models and found that a $^{12}$C$+^{16}$O reaction rate attenuation of 90\% matched best with observations. In contrast with all of the previously-discussed models, their simulations produced Ca-to-S mass ratios that are consistent with our results. 

\begin{figure*}[t]
\begin{center}
\textbf{Near-M$_{\rm Ch}$ Incompletely Shocked Nucleosynthesis Results} \\
\includegraphics[width=0.47\textwidth]{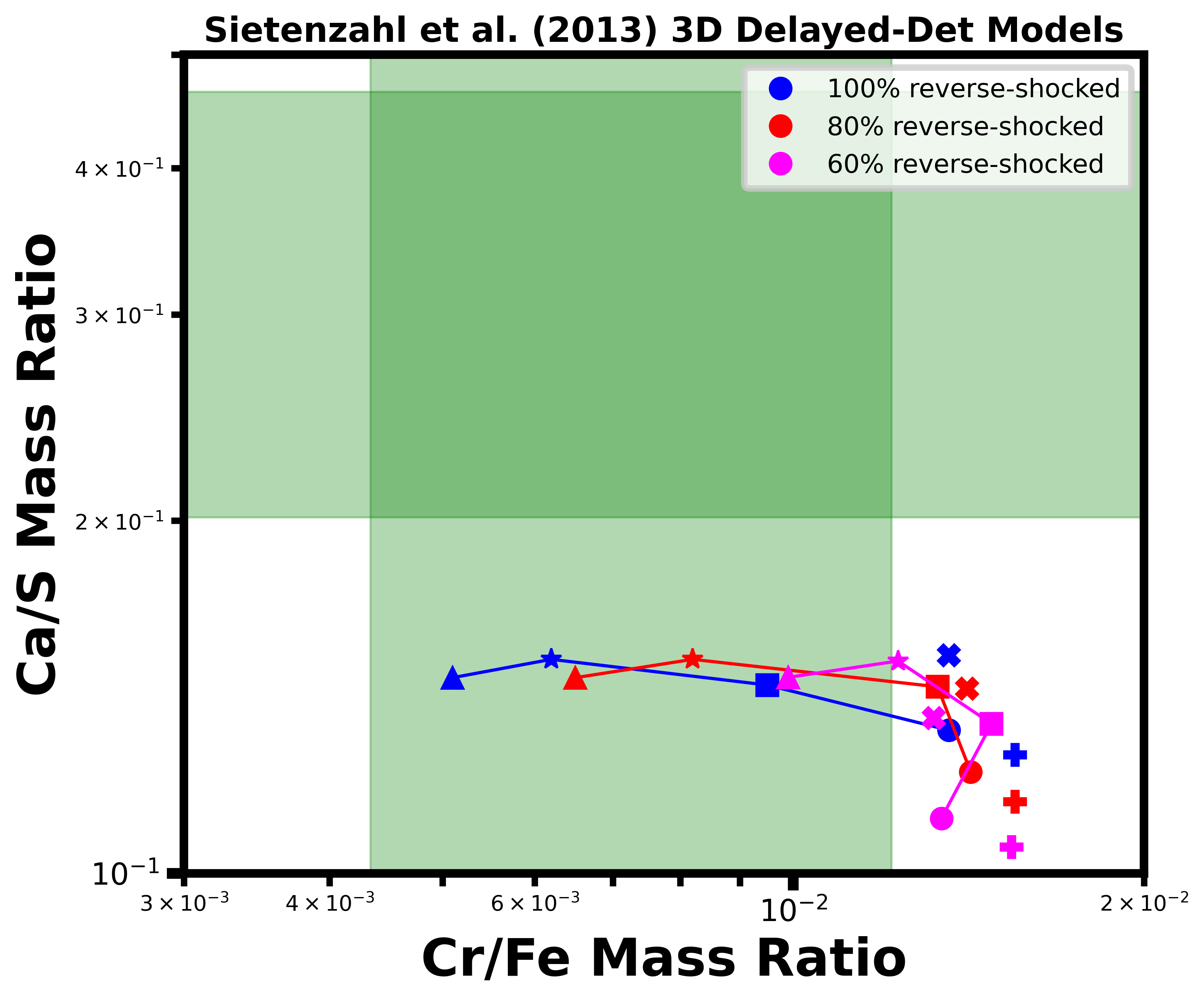}
\includegraphics[width=0.47\textwidth]{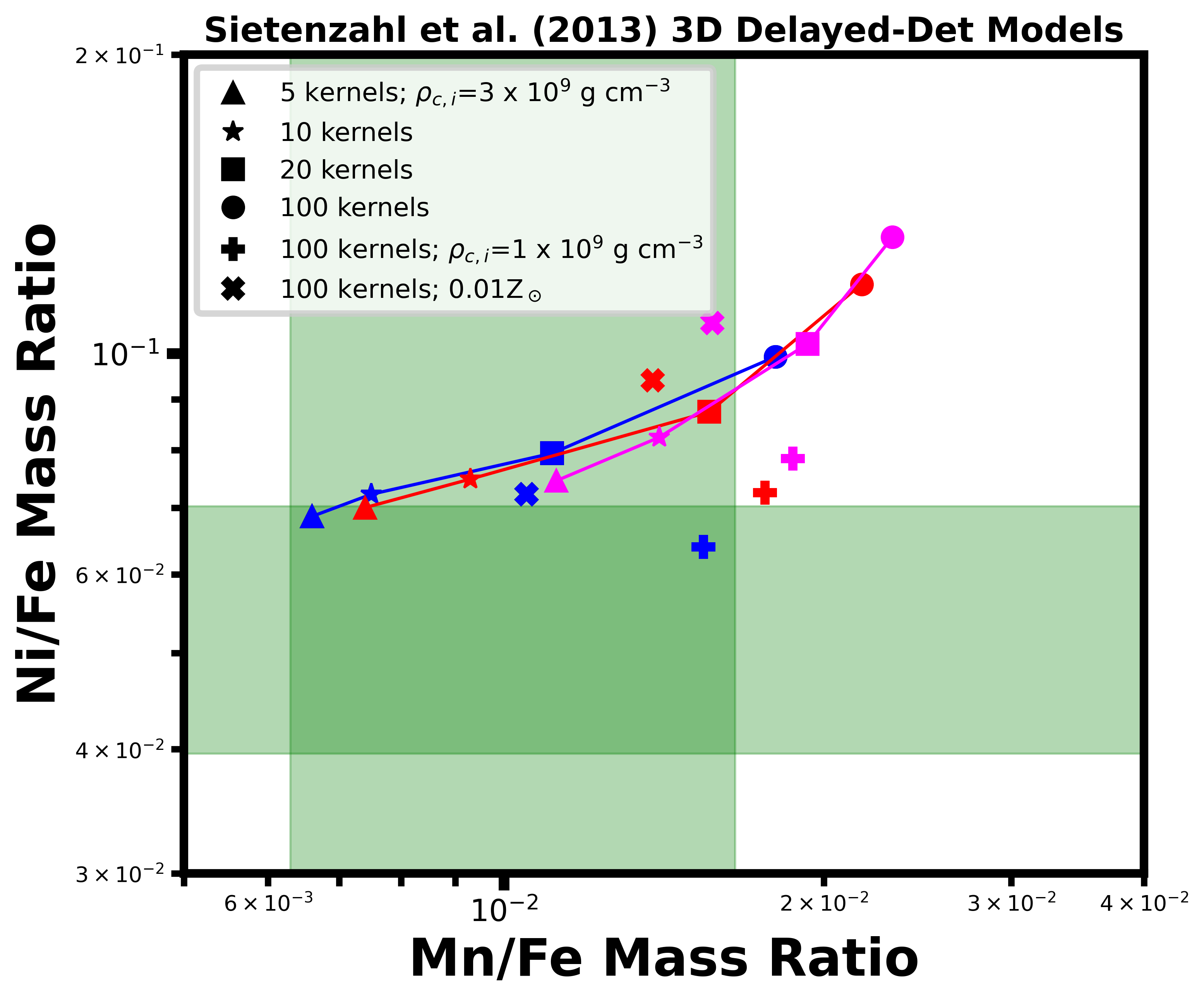} \\
\includegraphics[width=0.47\textwidth]{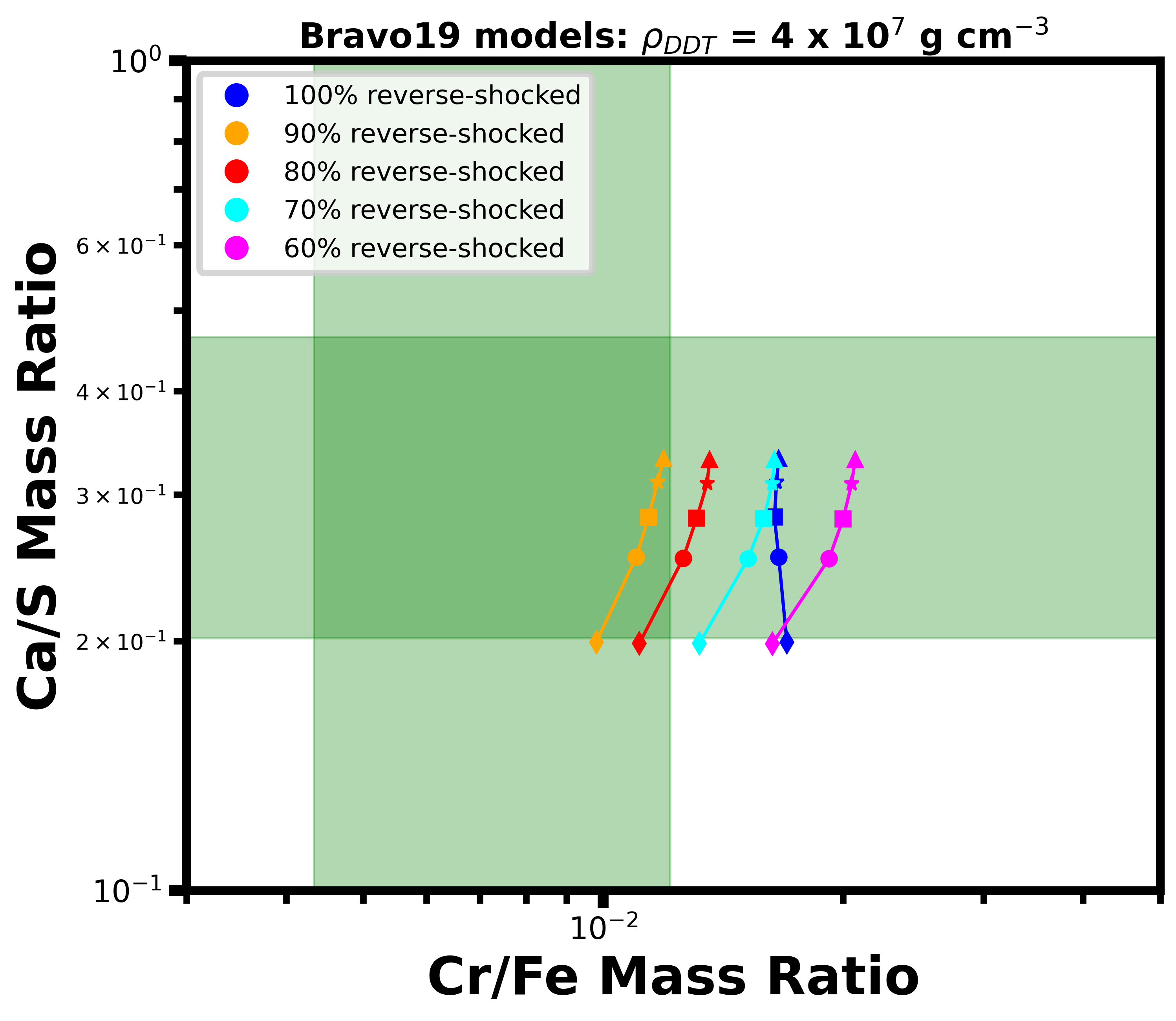}
\includegraphics[width=0.47\textwidth]{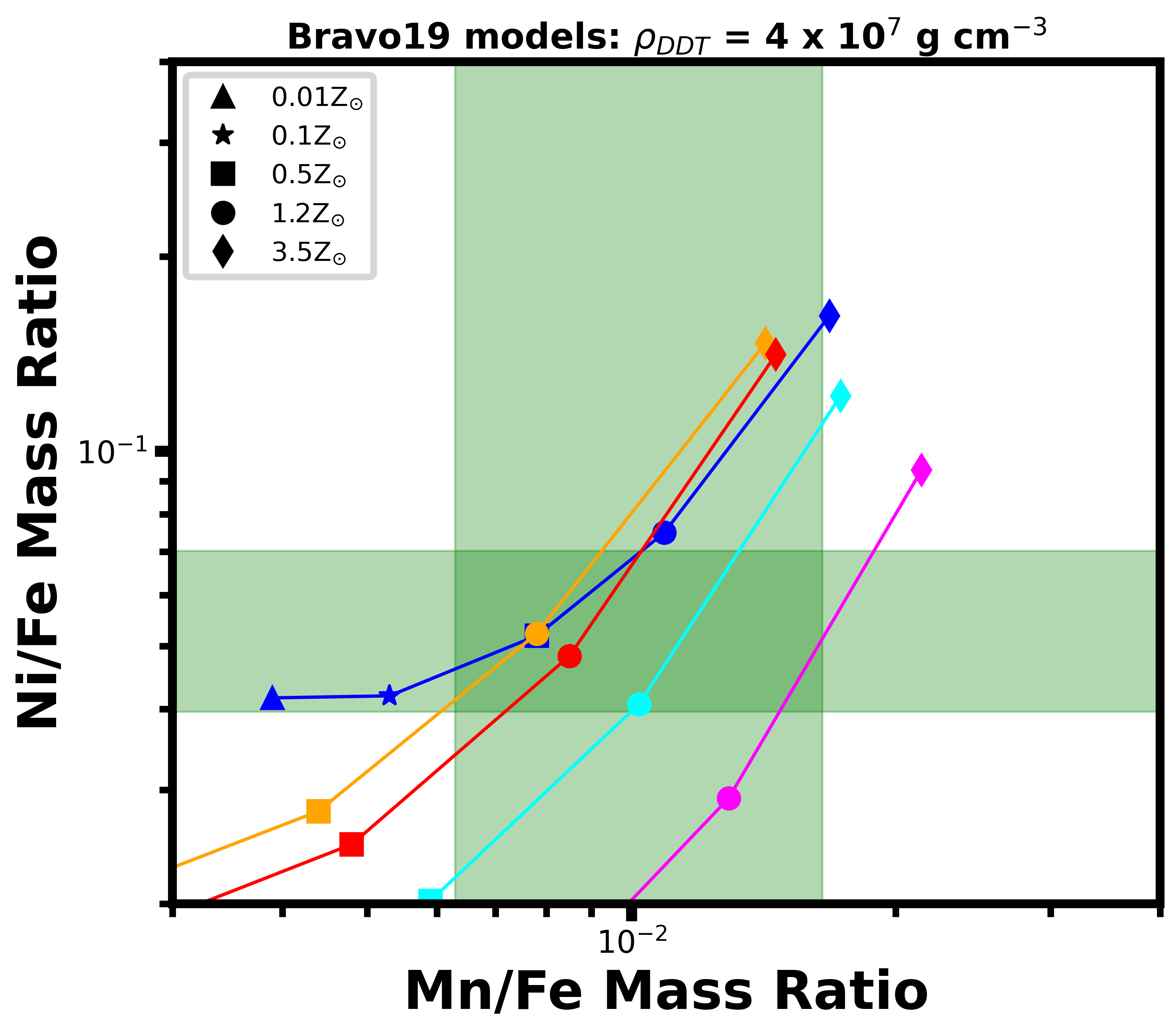}
\end{center}
\vspace{-5mm}
\caption{\footnotesize{A subset of near-M$_{\rm Ch}$ mass ratios from \cite{seitenzahl13} (top) and \cite{bravo19} (bottom) with differing amounts of shocked ejecta. For plotting clarity, we include fewer parameter combinations and step sizes of 20\% for the S13 data. For the B19 models, we only plot models with $\rho_{\rm DDT}=4 \times 10^7$ g cm$^{-3}$ as no others matched our estimates. Each row of plots is a pair, with labels split between the two plots.}}
\label{fig:nearCh_shockedperc}
\end{figure*}

Our estimates favor models from \cite{bravo19} with higher deflagration-to-detonation transition densities ($\sim$4 $\times 10^7$ g cm$^{-3}$) and moderate metallicities (Z=0.3--1.0Z$_\odot$). However, while their Mn/Fe and Ni/Fe predictions are consistent with our estimates, their predicted Cr/Fe mass ratios differ from our estimate by $\gtrsim$3-$\sigma$. When we also consider the nucleosynthesis dependencies of other papers, we suggest that fewer ignition kernels and/or a lower initial WD density might help to remove this discrepancy in Cr/Fe mass ratio. Such changes would likely then require a higher progenitor metallicity to raise the Mn/Fe and Ni/Fe mass ratios back to our estimated values.

\subsubsection{Incomplete reverse-shock propagation in near-M$_{\rm Ch}$ models}
As mentioned in Section~\ref{subsec:unshocked_obs}, we expect roughly 10--40\% of the ejecta in Kepler's SNR to be unshocked and thus not emitting in X-Rays. To account for this, we obtained Lagrange mass coordinate data from the simulation papers mentioned above and extracted simulated mass ratios from material present in the outer 60--100\% of ejecta. Figure~\ref{fig:nearCh_shockedperc} shows a subset of near-M$_{\rm Ch}$ models from \cite{seitenzahl13,bravo19} evaluated at various percentages of reverse-shocked ejecta. See Appendix~\ref{appen:comprehensive} for comprehensive plots detailing all model \& shocked-ejecta percentage combinations.

For the \cite{seitenzahl13} 3D delayed-detonation models, excluding interior ejecta resulted in larger IGE/Fe mass ratios. Given that their 100\% shocked models already over-predicted IGE/Fe ratios, unshocked ejecta fractions of $\gtrsim$20\% are strongly disfavored, even for low density and/or low metallicity progenitors.

For the \cite{bravo19} SNR-calibrated models, an unshocked fraction of 5--20\% produced Cr/Fe mass ratios that were more consistent with our observations than from fully shocked ejecta. These models still favor high deflagration-to-detonation densities but imply higher Z=1--2Z$_\odot$ progenitor metallicities instead of the $\sim$0.3--1Z$_\odot$ of fully shocked models.

We were unable to obtain Lagrange mass coordinate data for the updated W7, WDD2, and \cite{leung18} models, but analyses of other models showed that observed IGE/Fe ratios generally increase with less complete propagation of the RS. Thus, these models are likely even less consistent with our estimates when considering incomplete reverse-shock propagation.

\subsubsection{Near-M$_{\rm Ch}$ Model Summary}

\begin{figure*}
\begin{center}
\textbf{Sub-M$_{\rm Ch}$ Type Ia Model Nucleosynthesis Results}\\
\includegraphics[width=0.95\textwidth]{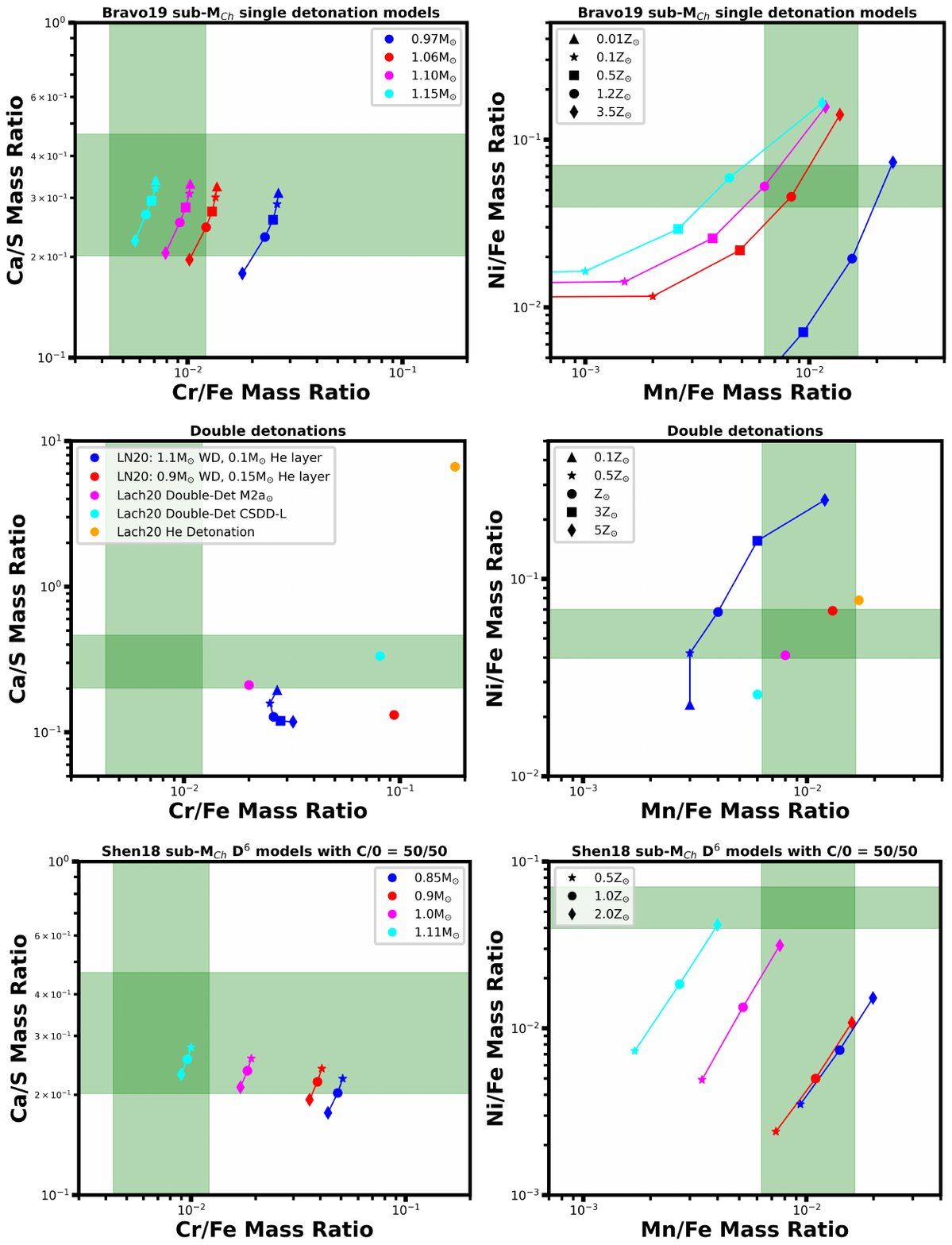}
\end{center}
\vspace{-5mm}
\caption{\footnotesize{Our estimated mass ratios for ejecta in Kepler's SNR (green bars, representing 90\% confidence intervals) compared to the results of various sub-M$_{\rm Ch}$ models. Bravo19 refers to \citep{bravo19}; LN20 refers to \cite{leung20}; Lach20 refers to \cite{lach20}, and Shen18 refers to \cite{shen18b}. Each row of plots is a pair, with labels split between the two plots.}}
\label{fig:SubCh_simulations}
\end{figure*}

In summary, only a subset of the models reported by \cite{bravo19} directly match with our observations. However, certain uninvestigated progenitor and explosion property parameter combinations for other simulations (e.g., \citealt{seitenzahl13}) show promise. Models with fewer ignition kernels (and thus higher $^{56}$Ni production), high deflagration-to-detonation transition densities, moderate metallicities, and shocked ejecta percentages of $\sim$80\% are favored.

\begin{figure*}
\begin{center}
\textbf{Sub-M$_{\rm Ch}$ Incompletely Shocked Nucleosynthesis Results} \\
\includegraphics[width=0.47\textwidth]{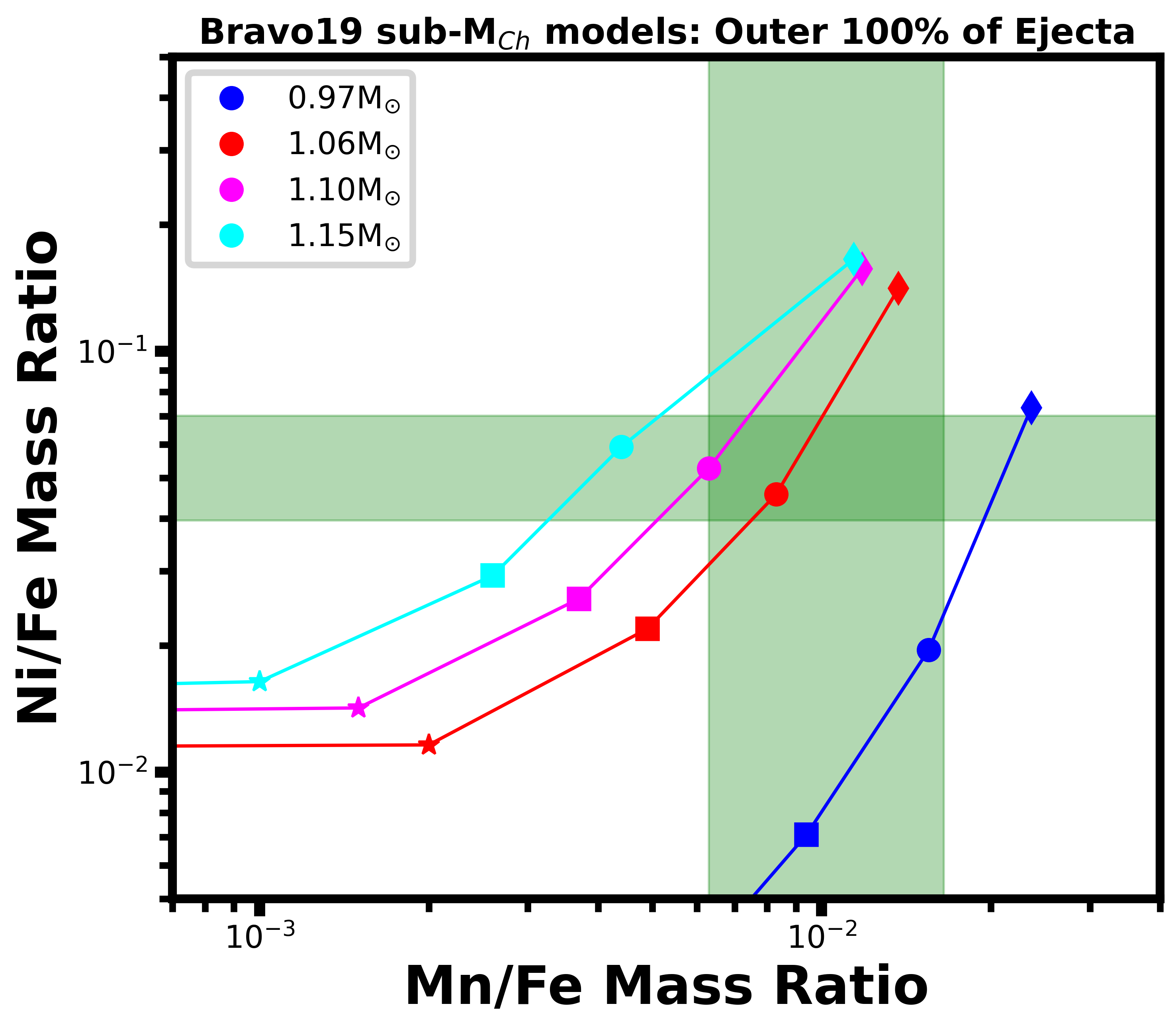}
\includegraphics[width=0.47\textwidth]{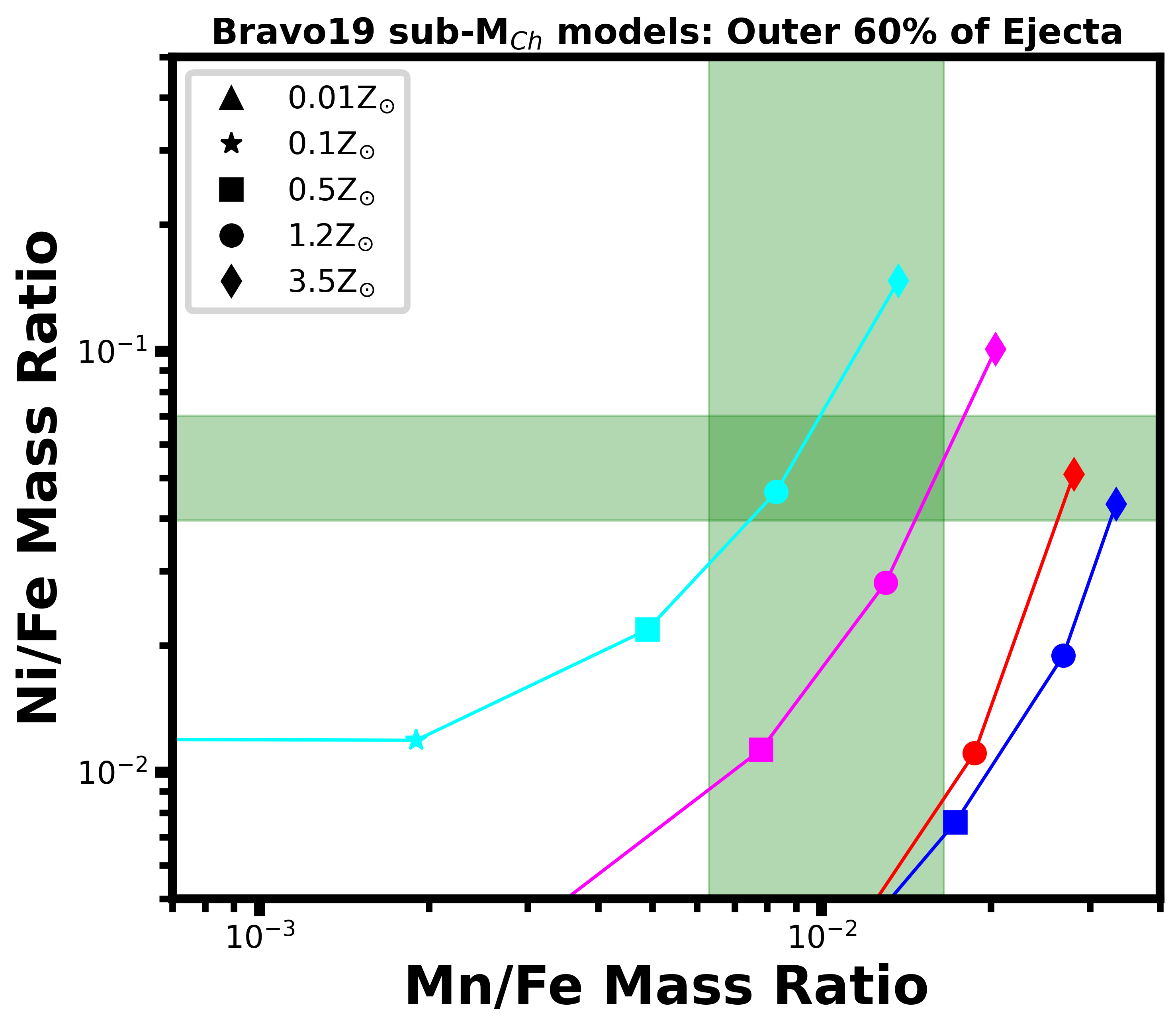}
\includegraphics[width=0.47\textwidth]{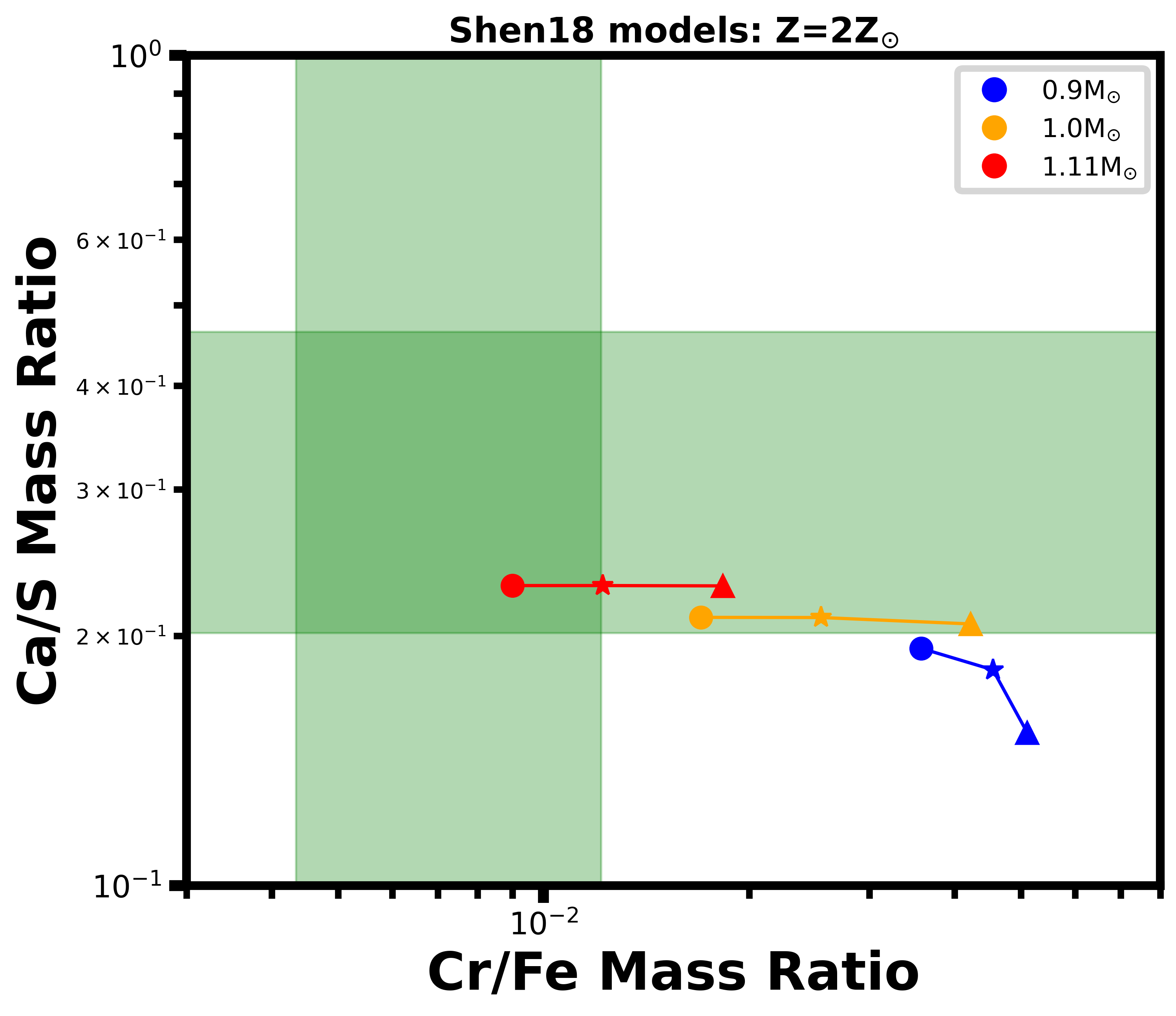}
\includegraphics[width=0.47\textwidth]{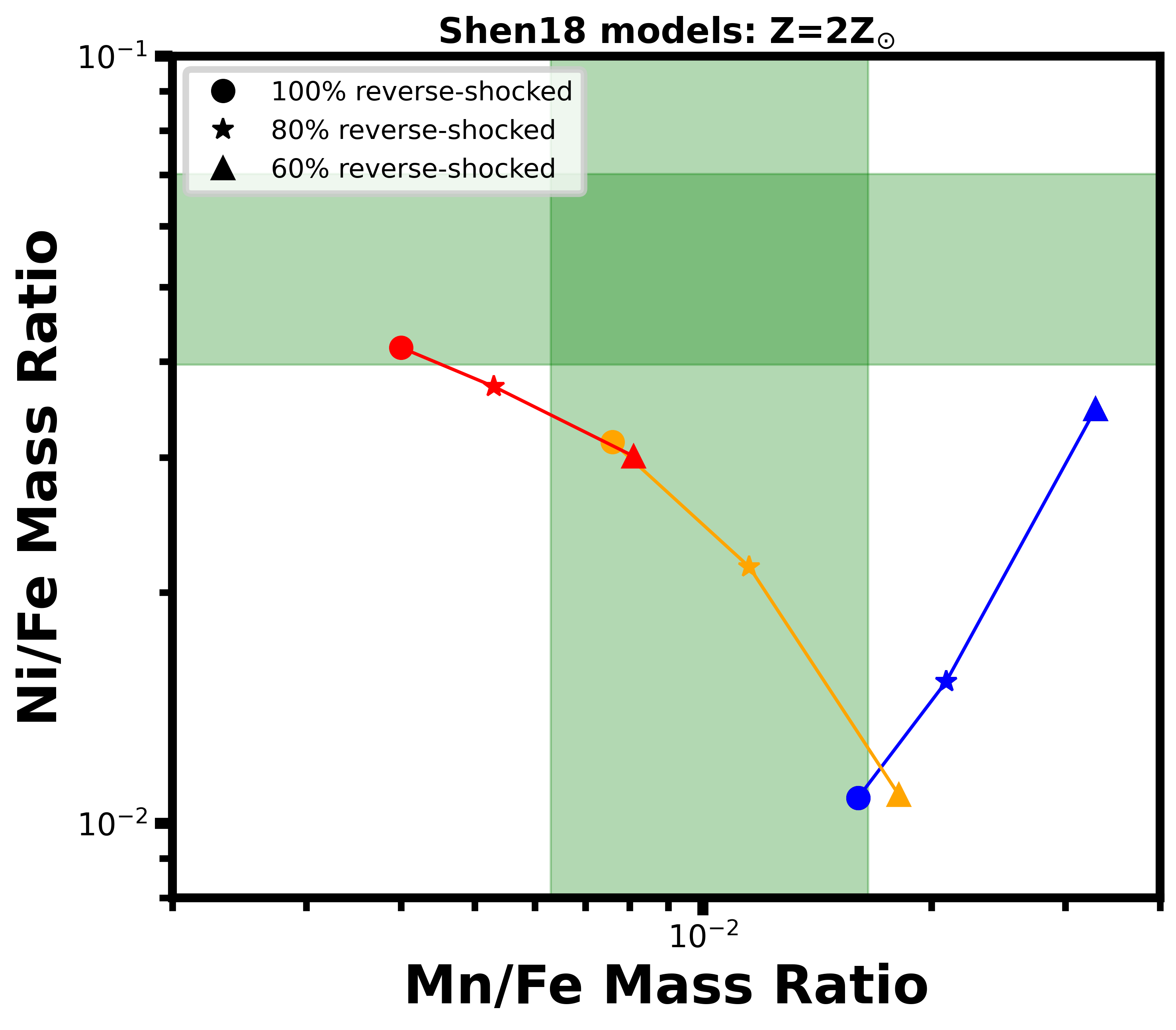}
\end{center}
\vspace{-5mm}
\caption{\footnotesize{(Top): mass ratios from \cite{bravo19} evaluated for the outer 100\% and 60\% of ejecta. The 70--90\% reverse-shocked values lie between the points plotted, and the Ca/S \& Cr/Fe mass ratios support the same progenitors. (Bottom): mass ratios from \cite{shen18b}, evaluated for differing amounts of shocked ejecta and Z=2Z$_\odot$ as no other metallicities matched our estimates. Each row of plots is a pair, with labels split between the two plots.}}
\label{fig:subCh_shockedperc}
\end{figure*}

Importantly, an attenuation rate of 90\% for the $^{12}$C$+^{16}$O reaction rate seems necessary to match our estimated Ca/S mass ratio in Kepler's SNR. Without this detail, models systematically underproduce Ca/S. However, few multi-dimensional models have been run using this attenuated rate. 

\subsection{Sub-M$_{\rm Ch}$ Explosion Models}
Common sub-M$_{\rm Ch}$ models include single detonations (the ``classic'' mergers), stable double detonations (DDet), and dynamically unstable double detonations (the D$^6$ model). In Figure~\ref{fig:SubCh_simulations}, we present our mass ratio estimates compared to the nucleosynthesis yields of sub-M$_{\rm Ch}$ simulations.

The top row of Figure~\ref{fig:SubCh_simulations} presents mass ratios from the sub-M$_{\rm Ch}$ models of \cite{bravo19} using the 90\% attenuated $^{12}$C$+^{16}$O reaction rate. These Cr/Fe mass ratios are much closer to our estimates than those from their near-M$_{\rm Ch}$ models. Extrapolating between their sub-M$_{\rm Ch}$ models, a primary mass around 1.07M$_\odot$ and metallicity of $\sim$1--2.5Z$_\odot$ produces Ca/S and IGE/Fe ratios that are most consistent with our estimates. 

The middle row of Figure~\ref{fig:SubCh_simulations} presents the mass ratios from the 2D models of \cite{leung20} and the 3D models of \cite{lach20}, both of whom investigated stable double-detonations of sub-M$_{\rm Ch}$ WDs with thick He layers. Interestingly, \cite{lach20} found that most of the $^{55}$Mn was synthesized in the helium detonation---not the WD core---and dominates the total nucleosynthesis when the He-shell to WD mass ratio is relatively large. The results of \cite{leung20} are consistent with this finding; at solar metallicity, their model with a higher fraction of He produces significantly more Mn (the red vs the blue circle data points). However, all of these models vastly overproduce Cr compared to our estimated mass ratios.

The bottom row of Figure~\ref{fig:SubCh_simulations} presents results of \cite{shen18b}. They modeled spherically symmetric simulations of naked sub-M$_{\rm Ch}$ ($\sim$0.8--1.1M$_\odot$) CO WD detonations, used as a proxy for the D$^6$ method assuming a very thin He layer. They investigated the effects of metallicity, primary WD mass, initial WD C/O mass ratio, attenuation of the $^{12}$C$+^{16}$O reaction rate, and the effect of incomplete reverse-shock propagation on observed mass ratios (see Figures~9 and 10 of their paper). We only include their C/O=50/50 results as their C/O=30/70 models were less consistent with our estimates. Their results favor high metallicities of Z$\gtrsim$2Z$_\odot$, progenitor masses of $\sim$1.1M$_\odot$, and a 90\% attenuation in the C+O reaction rate.

\subsubsection{Effects of incomplete reverse-shock propagation}
Figure~\ref{fig:subCh_shockedperc} shows a subset of sub-M$_{\rm Ch}$ models---from \cite{bravo19,shen18b}---evaluated at various fractions of shocked ejecta. See Appendix~\ref{appen:comprehensive} for more comprehensive plots.

For the \cite{bravo19} SNR-calibrated sub-M$_{\rm Ch}$ models, greater amounts of unshocked ejecta resulted in larger Cr/Fe and Mn/Fe mass ratios. These models still favor progenitor metallicities of Z=1--2Z$_\odot$, but greater unshocked material allows for progenitor masses of up to $\sim$1.2M$_\odot$ compared to the 100\% reverse-shocked model prediction of $\sim$1.07M$_\odot$.

For the \cite{shen18b} D$^6$ model, greater amounts of unshocked interior ejecta resulted in larger Cr/Fe and Mn/Fe mass ratios. The Cr/Fe mass ratio becomes inconsistent with our models for unshocked percentages $\gtrsim$20\%. However, even then, either the Ni/Fe or Mn/Fe mass ratios are still inconsistent with our 90\% confidence intervals. Of particular note is that the Ni/Fe mass ratio increases with more unshocked ejecta for higher mass progenitors, but decreases for lower-mass progenitors.

We could not obtain precise data from the 2D Double-Detonation modes of \cite{leung20,lach20}, but other models showed that IGE/Fe ratios generally increase with greater amounts of unshocked ejecta. This is consistent with \cite{lach20} who report that Mn/Fe ratios are higher in the outer layers of the SNR. As all the fully shocked DDet models already overproduced Cr/Fe by a factor of $\gtrsim$2, we can conclude that the stable DDet origin scenario remains inconsistent with Kepler's SNR even accounting for unshocked ejecta.

\subsubsection{Sub-M$_{\rm Ch}$ Model Review}
In summary, the \cite{bravo19} single-detonation and the \cite{shen18b} D$^6$ model can match our estimated mass ratios---favoring high progenitor metallicities ($\sim$2Z$_\odot$) and masses ($\sim$1.1M$_\odot$), while the stable DDet models of \cite{lach20,leung20} do not match our results. However, the D$^6$ model of \cite{shen18b} is only consistent with our results if we extrapolate their model to metallicities Z$\gtrsim$2Z$_\odot$. Similar to the near-M$_{\rm Ch}$ results, an attenuation rate of 90\% for the $^{12}$C$+^{16}$O reaction rate is necessary to match our estimated Ar/S and Ca/S mass ratios in Kepler's SNR.

\section{Conclusions}
\label{sec:conc}

In this paper, we have analyzed the entirety of Kepler's SNR, fitting its 0.6--8.0~keV spectra with plasma models in order to measure the mass ratios of X-Ray emitting ejecta. We have found that relatively simple models can produce reasonable-looking fits to this incredibly complex spectra, although with high reduced-$\chi^2$ values of $\sim$2.5--3.5. However, these residuals can be entirely explained by unaccounted-for telescope systematics: e.g., the 5--20\% effective area calibration uncertainties in {\it Suzaku} XIS detectors. This can dominate over the statistical photon uncertainty for high signal-to-noise spectra.

To properly correct for these energy-dependent uncertainties, we generated 100 mock effective area curves and used MCMC-based fitting to analyze the resulting 100 spectra, characterizing the resulting variance in best-fit parameter spread as the telescope calibration uncertainty. 

Additionally, we investigated the effects of the different filling factor assumptions, necessary when element abundances are allowed to vary between multiple plasma components. This unknown can introduce up to a $\sim$30\% uncertainty on mass ratios estimated from multi-component models, larger than the uncertainty propagated from spectral fits high signal-to-noise X-Ray data.

The systematic uncertainties investigated in this paper will be extremely useful for future studies of global SNR spectra, especially those with incredibly high signal-to-noise: e.g., Cassiopeia~A and Tycho's SNR. Analyzing the broad spectra of the entire SNRs is essential to calculating total mass ratios, as analyzing small regions does not provide a full picture of the SNR, doing a grid based analysis introduces countless filling factor assumptions, and analyzing only restricted bandpasses can prevent accurate identification of all the X-Ray emitting plasma components.

\subsection{Constraints on Kepler's Progenitor}
Our estimated mass ratios for Kepler's SNR are broadly consistent with past observational works on ejecta mass ratios in Kepler's SNR. Notably, our reported mass ratio uncertainties are of a similar magnitude or much smaller than previous works, even though we are including two additional source of error: {\it Suzaku} effective area uncertainties and the uncertainty associated with various filling factor assumptions. This indicates that our use of multiple components across a wide bandpass and the full SNR helped to better constrain the various plasma properties. 

We make the following conclusions and associations about the progenitor of Kepler's SNR in order of significance:
\begin{enumerate}
    \item \textbf{Our strongest conclusion is that our observed Ca/S mass ratio estimate requires a 90\% attenuation in the standard $^{12}$C$+^{16}$O reaction rate} (e.g., \citealt{mr17,shen18b,bravo19}. 
    \item \textbf{A single-detonation sub-M$_{\rm Ch}$ explosion with a primary mass of 1.07--1.15M$_\odot$, metallicity of 1--2.5Z$_\odot$, and shocked ejecta percentages of 60--100\% produces mass ratios within 2-$\sigma$ of all of our estimates.} This 1D model from \cite{bravo19} uses the 90\% attenuated C$+$O reaction rate and should be investigated in 3D.
    \item \textbf{Dynamically stable double-detonation models (involving a thick He layer) are disfavored} as they vastly overproduce Cr relative to Fe.
    \item \textbf{Our results are possibly consistent with a delayed detonation near-$M_{\rm Ch}$ origin.} One 1D model from \cite{bravo19} with $\rho_{DDT} \approx 4\times 10^{7}$ g cm$^{-3}$ and Z=1--2Z$_\odot$ matches with our results for a narrow range of shocked ejecta---80--95\%. Alternatively, the \cite{seitenzahl13} 3D models might produce consistent results if the simulation used the 90\% attenuated C$+$O reaction rate and a parameter combination involving relatively few ignition kernels, low central density ($\rho_c$$\approx $1--3$\times 10^9$ g cm$^{-3}$), $\rho_{DDT}$$\approx $2--4 $\times 10^{7}$ g cm$^{-3}$, and Z$\approx$0.2--1.5Z$_\odot$.
    \item \textbf{The D$^6$ explosion method might be consistent with our mass ratio estimates, but only for the most massive progenitors ($\gtrsim$1.1M$_\odot$), high metallicities (Z$\gtrsim$2Z$_\odot$), and requiring $\lesssim$10\% of the ejecta in Kepler's SNR is unshocked.} As other sources suggest that Kepler's SNR has 10--40\% unshocked material, this scenario is only barely consistent.
\end{enumerate}

\acknowledgments

We thank Dr. Adam Foster for the many discussions about AtomDB and for providing updated Fe emission line data. We thank Dr. Keith Arnaud for providing significant Xspec expertise: suggestions, clarifications, and updates to the software. We thank Drs. Sylvain Korzennik, Craig Gordon, and Josh Wing for helping us to get Xspec working on the Smithsonian Institution High Performance Cluster. We thank Drs. Eduardo Bravo, Carles Badenes, Ivo Seitenzahl, and Ken Shen for helpfully providing Lagrange Mass Coordinate nucleosynthesis tables from their simulations' results.

Dr. Pat Slane acknowledges support from NASA Contract NAS8-03060. LAL acknowledges support by the Simons Foundation, the Heising-Simons Foundation, and a Cottrell Scholar Award from the Research Corporation for Science Advancement. This work was performed in part at the Simons Foundation Flatiron Institute's Center for Computational Astrophysics during LAL's tenure as an IDEA Scholar. 

This research made use of the data analysis software: HEASoft (6.30, \small http://heasarc.gsfc.nasa.gov/ftools, \normalsize [ascl:1408.004]), XSPEC (v12.12.0; \citealt{arnaud96}), AtomDB (v3.0.10 \citealt{smith01,foster12}), PyAtomDB (\small https://atomdb.readthedocs.io/en/master/), \normalsize and CALDB (\small https://heasarc.gsfc.nasa.gov/FTP/caldb, \normalsize XIS 20160607). This research made use of data obtained from the {\it Suzaku} satellite, a collaborative mission between the space agencies of Japan (JAXA) and the USA (NASA). The computations in this paper were conducted on the Smithsonian High Performance Cluster (SI/HPC), Smithsonian Institution. https://doi.org/10.25572/SIHPC. This work made use of the Heidelberg Supernova Model Archive (HESMA), https://hesma.h-its.org.

\appendix

\section{Element Mass Ratio Calculations}
\label{appen:MRs}
Our spectral models are composed of 5 plasma components: one shocked CSM/ISM \texttt{vpshock} component, two shocked IME-dominated ejecta \texttt{vpshock} components, one shocked Fe-dominated ejecta component \texttt{vvnei}, and one non-thermal emission component \texttt{srcut}. To calculate ejecta mass ratios, we consider the following: 

\begin{deluxetable}{lccc}[!h]
\tablecolumns{4}
\tablewidth{0pt} 
\tablecaption{Final Model Parameters \label{table:fitvals}} 
\tablehead{ \colhead{Parameter} & \colhead{Value} &\colhead{$\sigma_{\rm {cal}}$\tablenotemark{a}} & \colhead{$\overline{\sigma_{\rm {stat}}}$\tablenotemark{a}} }
\startdata
$N_{\rm{H}}$ (10$^{22}$ cm$^{-2}$)  & 0.763 $\pm$ 0.051 & 0.036 & 0.037 \\ \hline
CSM Component: \texttt{vpshock}\\
$kT_{\rm e}$ (keV)   & 0.405 $\pm$ 0.157 & 0.130 & 0.087  \\
Abundance\tablenotemark{b} (solar) N  & 3\tablenotemark{c} & -- & -- \\
O  & 0.59 $\pm$ 0.25 & 0.23& 0.10\\
Si  & 0.41 $\pm$ 0.25 & 0.19& 0.16\\
Fe  & 0.59 $\pm$ 0.26 & 0.22& 0.13\\
Ionization Timescale ($10^{11}$ cm$^{-3}$ s)  & 4.36 $\pm$ 2.04& 1.70 & 1.13 \\
Normalization	(cm$^{-5}$)   & 0.173 $\pm$ 0.107 & 0.091 & 0.056\\\hline
Ejecta 1 component: \texttt{vpshock}\\
$kT_{\rm e}$ (keV)    & 0.606 $\pm$  0.124& 0.084 & 0.092 \\
Abundance\tablenotemark{b} (10${^5}$ solar) O & 0.118 $\pm$ 0.050 & 0.042 & 0.028\\
Mg  & 0.104 $\pm$  0.037& 0.034& 0.015 \\
Ne  & 0.088 $\pm$  0.097&0.086& 0.043\\
Si  & 1\tablenotemark{c}& --& --\\
S  & 1.06 $\pm$  0.13& 0.13& 0.042\\
Ar  & 1.35 $\pm$ 0.40& 0.38& 0.12\\
Ca  & 2.20 $\pm$ 0.63& 0.59& 0.22\\
Fe  & 0.75 $\pm$ 0.33& 0.25 & 0.22\\
Ionization Timescale($10^{11}$ cm$^{-3}$ s) & 2.52 $\pm$  1.43& 1.26& 0.69\\
Redshift ($10^{-3}$)  & -0.561 $\pm$  0.128& 1.14& 0.56\\
Normalization	(10${-6}$ cm$^{-5}$)   & 2.13 $\pm$ 0.84& 0.75 & 0.38\\\hline
Ejecta 2 component: \texttt{vpshock}\\
$kT_{\rm e}$ (keV)   & 1.71 $\pm$  0.20& 0.175 & 0.100 \\
Abundance\tablenotemark{b} (10$^5$ solar) Fe  & 0.25 $\pm$  0.17& 0.09& 0.14\\
Ionization Timescale ($10^{10}$ cm$^{-3}$ s)  & 5.86 $\pm$ 1.01 & 0.58& 0.83\\
Redshift ($10^{-3}$)   & -1.27 $\pm$ 0.73& 0.57& 0.45\\
Line Broadening (E/6 keV)  & 0.015 $\pm$ 0.002& 0.0012& 0.0017\\
Normalization	(10${-6}$ cm$^{-5}$)  & 2.20 $\pm$  0.54& 0.46& 0.29\\\hline
Ejecta 3 component: \texttt{vvnei}\\
$kT_{\rm e}$ (keV)   & 5.96 $\pm$  0.80 & 0.68& 0.43\\
Abundance\tablenotemark{b} (10${^5}$ solar) Cr  & 0.965 $\pm$ 0.225 & 0.104& 0.200\\
Mn  & 1.87$\pm$ 0.39 & 0.12& 0.37\\
Fe  & 1\tablenotemark{c} & --& --\\
Ni  & 1.35$\pm$ 0.19& 0.08& 0.17 \\
Ionization Timescale ($10^{9}$ cm$^{-3}$ s)  & 3.7\tablenotemark{c}& --& --\\
Redshift ($10^{-3}$)  & -2.43 $\pm$ 0.77& 0.15& 0.76\\
Line Broadening (E/6 keV)  & 0.72\tablenotemark{c} & -- & --\\
Normalization	(10${-6}$ cm$^{-5}$)   & 2.06 $\pm$ 0.40& 0.33 & 0.24\\\hline
Synchrotron component: \texttt{srcut}\\
Radio Spectral Index $\alpha$  & -0.71\tablenotemark{c} & --& --\\
Break Frequency (10$^{17}$ Hz)  & 2.19 $\pm$ 0.33 & 2.92& 1.58\\
Normalization (1 GHz flux; Jy)   & 25.6 $\pm$ 3.5 & 3.0 & 1.9 \\
\enddata
\vspace{-1mm}
\tablenotetext{a}{\footnotesize{$\sigma_{\rm {cal}}$ and {$\overline{\sigma_{\rm {stat}}}$} represent the spread in the 100 MCMC parameter means and the average parameter spread in each MCMC run, respectively. }}
\vspace{-1mm}
\tablenotetext{b}{\footnotesize{Unmentioned element abundances are fixed to solar. Ejecta 2 element abundances other than Fe were linked to those in Ejecta 1.}}
\vspace{-1mm}
\tablenotetext{c}{\footnotesize{These values were frozen during fitting. Thawing the ionization timescale and line broadening parameters in Ejecta 3 resulted in fits not properly capturing Fe-K emission. Additionally, the line broadening parameters of the CSM and Ejecta 1 components went to $\lesssim 10^{-5}$ during fitting and became independent of $\chi^2$ minimization. Thus, we fixed them to $10^{-5}$. Other parameters were frozen to values found in previous papers' as discussed in the text.}}
\end{deluxetable}

The abundance of an element X is a number density relative to the solar abundance of that element w.r.t. hydrogen
\[abund_X = \frac{[n_X/n_H]}{[n_X/n_H]_{\odot}} = \frac{[X/H]}{[X/H]_\odot}\]
where $n_X$ is the number density of that element. In our models, the solar values are taken from \cite{wilms00}. The normalization of a \texttt{vpshock} or \texttt{vvnei} Xspec component is defined as:
\[{\rm norm} = \bigg(\frac{10^{-14}}{4\pi D^2}\bigg)\int n_en_HdV_{\rm emit} \approx  \bigg(\frac{10^{-14}}{4\pi D^2}\bigg) n_en_HV_{\rm emit} \]
where D is the distance to the source in cm, $n_e$ is the electron density (cm$^{-3}$), $n_H$ is the hydrogen density (cm$^{-3}$), and V$_{\rm emit}$ is the emitting volume of the source. We have simplified the integral by assuming constant density over the given volume. Additionally, we can write the electron density $n_e$ in terms of $n_H$: $n_e = C_e n_H$ where  C$_e$ is a multiplicative factor that can be calculated by summing over the electron contributions from all ionized elements. We estimated each the average ionization level of each element by using the pyatomdb \texttt{apec} module \texttt{apec.return\_ionbal}, inputting the best-fit plasma temperature and ionization timescale.

We can also write V$_{\rm emit} = $V$_{\rm tot} \times$ f, where V$_{\rm tot}$ is the total volume of Kepler's SNR and f is the filling factor of the ejecta material in that volume. 
\[{\rm norm} = \bigg(\frac{10^{-14}}{4\pi D^2}\bigg) C_en_H^2V_{\rm tot} \times f \]
The total mass of an element X is its number density times atomic mass (m$_{\rm amu, X}$) times volume:
\[M_{X} = n_Xm_{\rm amu, X}V_{\rm emit} = abund_X \times n_H \times [X/H]_{\odot} \times {\rm m}_{\rm amu, X} \times V_{\rm tot} \times f\]
We can re-write the normalization equation to solve for n$_H$:
\[n_H = \bigg(\frac{norm \times 4\pi D^2}{10^{-14}C_eV_{\rm tot} \times f}\bigg)^{0.5} = A\bigg(\frac{norm}{C_eV_{\rm tot}}\bigg)^{0.5} f^{-0.5} \]
where we have combined constants; A=$(4\pi D^2\times10^{14})^{0.5}$. This gives us a final equation
\begin{equation}
    M_{X} = abund_X \times AC_e^{-0.5} \times [X/H]_{\odot} \times norm^{0.5} \times {\rm m}_{\rm amu, X} \times V_{\rm tot}^{0.5} \times f^{0.5}
\end{equation}
The final mass estimate for an element in a single plasma component depends on the square root of that unknown filling factor. We must add the contributions from each component to get total mass.
\[M_{X,tot} = M_{X,1} + M_{X,2} + M_{X,3} + ... M_{X,N} \]
We can rewrite the individual mass terms M$_{X,1}=$ M$_{\rm X,ej1}f_{\rm ej1}^{0.5}$, where M$_{\rm X,ej1}$ contains all of M$_{\rm X,1}$ except for the filling factor term, and use the fact that we only have three ejecta-dominated components.
\[ M_{\rm X, tot} = M_{\rm X, ej1}f_{\rm ej1}^{0.5} + M_{\rm X, ej2}f_{\rm ej2}^{0.5} + M_{\rm X, ej3}f_{\rm ej3}^{0.5} \]
\begin{equation}
    M_{\rm X, tot} = \big(M_{\rm X, ej1} + \alpha M_{\rm X, ej2} + \beta M_{\rm X, ej3}\big)f_{\rm ej1}^{0.5}
\end{equation} 
where $\alpha=(f_{\rm ej2}/f_{\rm ej1})^{0.5}$ and $\beta=(f_{\rm ej3}/f_{\rm ej1})^{0.5}$.

\section{Filling Factor Assumptions}
\label{appen:FillFs}
As mentioned in Section~\ref{subsec:MRCalcs}, we use four assumptions about filling factor relationships (i.e., $\alpha$ and $\beta$) to calculate total ejecta masses and mass ratios.

\begin{enumerate}
\item Equal Volume: each component has an equal filling factor (equal emitting volume). This is the simplest solution where we simply set the filling factors equal to each other.
\[ M_{\rm X, tot} = \big(M_{\rm X, ej1} + M_{\rm X, ej2} + M_{\rm X, ej3}\big)f_{\rm ej1}^{0.5} \]
\item Electron Pressure Equilibrium: P${_e}$=$n_ekT_e$ is constant between ejecta components, which simplifies to
\[n_{e,1}T_{e,1} = n_{e,2}T_{e,2} \]
for two components. Plasma electron temperature is a fit parameter of our Xspec models, and $n_e$ can be rewritten in terms of n$_H$: $n_e = C_e n_H$. Our equation is becomes
\[ C_{e,1}n_{H,1}T_{e,1} = C_{e,2}n_{H,2}T_{e,2}\]
Plug in our above equation for $n_H$ to get
\[ C_{e,1}A_1\bigg(\frac{norm_1}{V_{\rm tot, 1} \times f_{1}}\bigg)^{0.5}T_{e,1} = C_{e,2}A_2\bigg(\frac{norm_2}{V_{\rm tot, 2} \times f_{2}}\bigg)^{0.5}T_{e,2}\]
The total volumes of the 2 components are the entire volume of Kepler's SNR and thus equal, as are all parameters except for $C_e$, the normalization, and electron temperature. We simplify:
\[ C_{e,1}norm_1^{0.5}f_{1}^{-0.5}T_{e,1} = C_{e,2}norm_2^{0.5}f_{2}^{-0.5}T_{e,2}\]
\[ f_{2}^{0.5} = \frac{norm_2^{0.5} \times C_{e,2}T_{e,2}}{norm_1^{0.5} \times C_{e,1}T_{e,1}}f_{1}^{0.5} = \frac{(norm^{0.5} \times C_{e}T_{e})_2}{(norm^{0.5} \times C_{e}T_{e})_1} f_{1}^{0.5}\]
In the above equation, all variables except for the filling factors are produced by our fit models.

Alternatively, we can write this equation in terms of physical plasma properties. We know that n$_e \propto n_H \propto f^{-0.5}$. Thus, we can define $\eta_e$ as $n_e$ evaluated for a filling factor of 1: n$_e=\eta_e f^{0.5}$.
\[ f_{\rm ej2}^{0.5} = \frac{(\eta_{e}T_{e})_{\rm ej2}}{(\eta_{e}T_{e})_{\rm ej1}} f_{\rm ej1}^{0.5} \]

\item Linking plasmas to specific regions in the SNR, done in \cite{katsuda15}, who identified IME-dominated emission as coming from a shell region 0.85--0.97 times the radius of the forward shock (R$_{\rm FS}$) and Fe-dominated emission as coming from a shell region 0.7--0.85R$_{\rm FS}$. They assumed the two IME-dominated plasmas were in pressure equilibrium and together filled the entire outer shell, and the Fe-dominated plasma filled its entire shell: 
\[ f_{\rm ej1}+f_{\rm ej2}=(0.97^3-0.85^3) \text{ and } f_{\rm ej3}=(0.85^3-0.7^3) \]

For the two IME-dominated ejecta components, the individual filling factors are related using the same pressure equilibrium assumption as described for method \#2.

\item As above, but enforced pressure equilibrium between the two shells in addition to pressure equilibrium within the shells: P$_{\rm ej3}$=P$_{\rm ej1}$+P$_{\rm ej2}$=2P$_{\rm ej1}$. The relation between the IME-components remains the same as in method \#3, but for $f_{\rm ej3}$ we instead use
\[n_{e_3}T_{e,3} = 2n_{e_1}T_{e,1} \]
The total volumes of the shells differ, so when we substitute in our equation for n$_H$ we can't cancel those terms, resulting in the final equation
\[ f_{\rm ej3}^{0.5} = \frac{(\eta_{e}T_{e})_{\rm ej3}}{2(\eta_{e}T_{e})_{\rm ej1}}\bigg(\frac{V_3}{V_1} \bigg)^{0.5} f_{\rm ej1}^{0.5} \]
\[ f_{\rm ej3}^{0.5} = \frac{(\eta_{e}T_{e})_{\rm ej3}}{2(\eta_{e}T_{e})_{\rm ej1}}\bigg(\frac{0.97^3-0.85^3}{0.85^3-0.7^3} \bigg)^{0.5} f_{\rm ej1}^{0.5}\]
\[f_{\rm ej3}^{0.5}\approx 0.55\frac{(\eta_{e}T_{e})_{\rm ej3}}{(\eta_{e}T_{e})_{\rm ej1}} f_{\rm ej1}^{0.5}\]
\end{enumerate}
Importantly, in all the above cases, we still do not---cannot---know $f_{\rm ej1}$ with our current measurements. Without an additional assumption, we can only estimate mass ratios, not total masses.

\section{Comprehensive Ejecta Mass Ratios From Simulations}
\label{appen:comprehensive}

\begin{figure*}
\begin{center}
\textbf{Seitenzahl et al. (2013) 3D Nucleosynthesis Results from Outer X\% of Ejecta} \\
\includegraphics[width=0.95\textwidth]{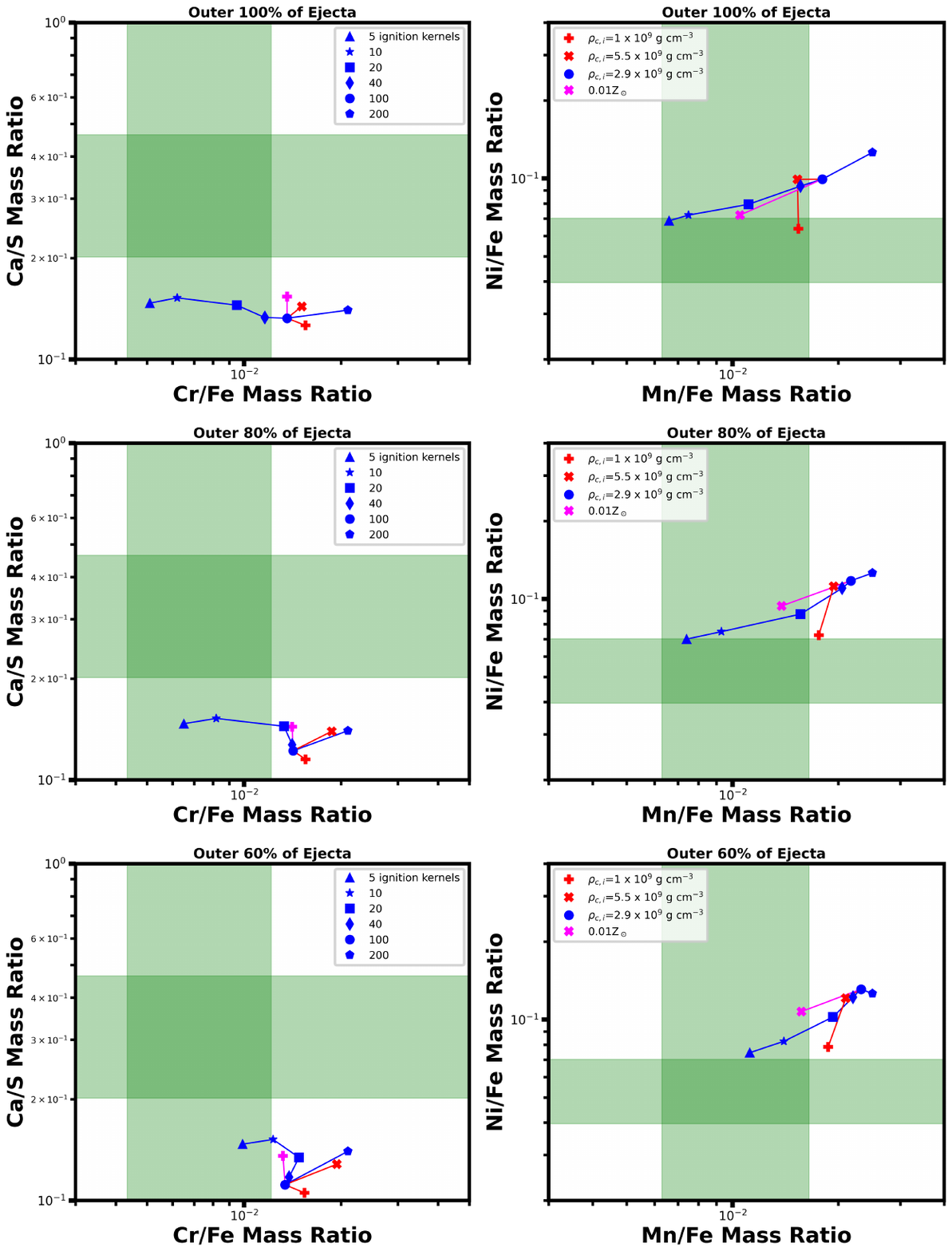}
\end{center}
\vspace{-5mm}
\caption{\footnotesize{Our estimated mass ratios for ejecta in Kepler's SNR (green bars, representing 90\% confidence intervals) compared to the results of \cite{seitenzahl13} evaluated at different percentages of shocked (and thus visible) ejecta. Each row of plots is a pair, with labels split between the two plots.}}
\label{fig:S13_all}
\end{figure*}

\begin{figure*}
\begin{center}
\textbf{Bravo et al. (2019) 1D, near-M$_{\rm Ch}$ SNR-Calibrated Nucleosynthesis from Outer X\% of Ejecta} \\
\includegraphics[width=0.95\textwidth]{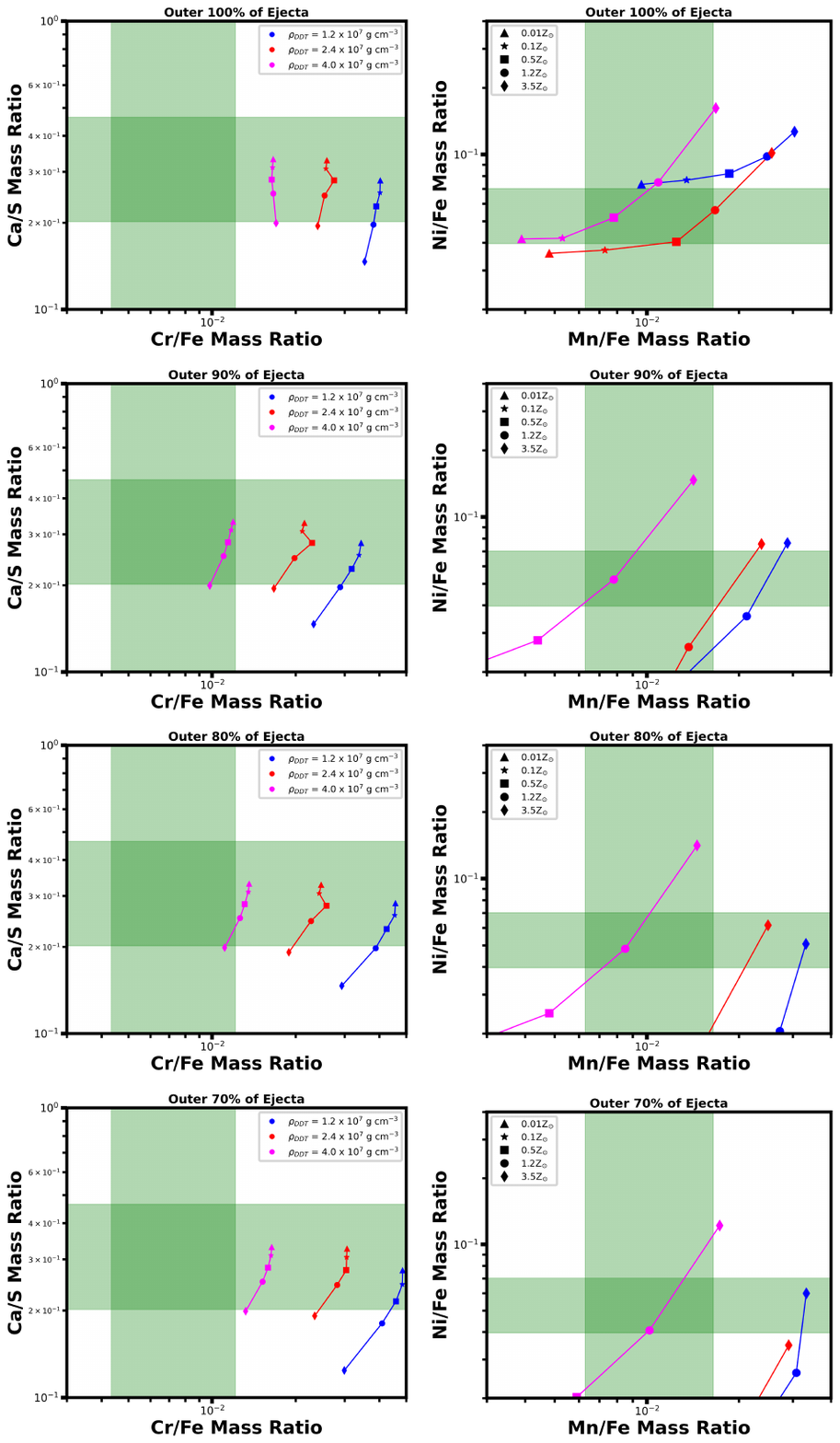}
\end{center}
\vspace{-5mm}
\caption{\footnotesize{Our estimated mass ratios for ejecta in Kepler's SNR (green bars, representing 90\% confidence intervals) compared to the near-M$_{\rm Ch}$ results of \cite{bravo19} evaluated at different percentages of shocked (and thus visible) ejecta. Each row of plots is a pair, with labels split between the two plots.}}
\label{fig:B19_all}
\end{figure*}

\begin{figure*}
\begin{center}
\textbf{Bravo et al. (2019) 1D sub-M$_{\rm Ch}$ Nucleosynthesis from Outer X\% of Ejecta} \\
\includegraphics[width=0.95\textwidth]{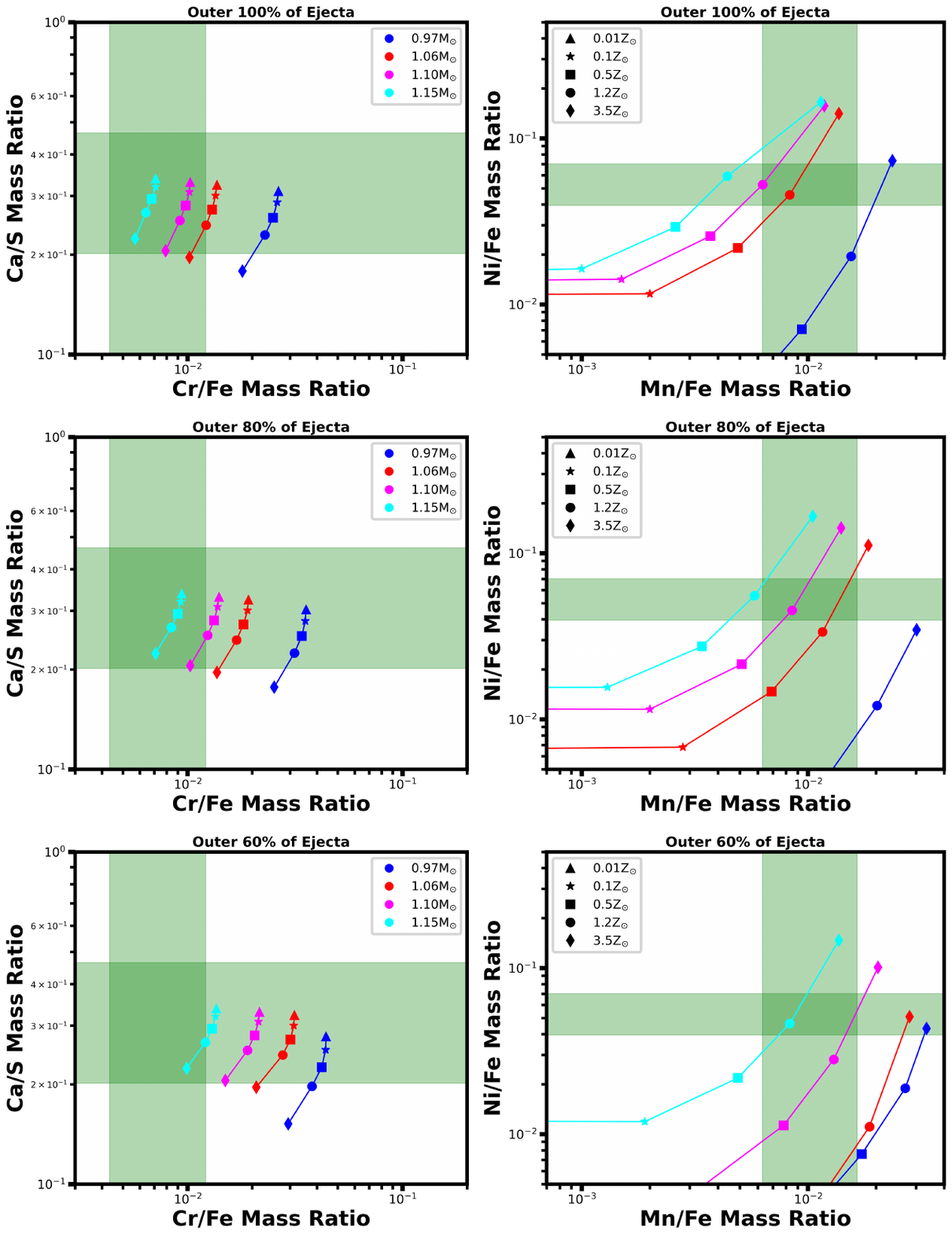}
\end{center}
\vspace{-5mm}
\caption{\footnotesize{Our estimated mass ratios for ejecta in Kepler's SNR (green bars, representing 90\% confidence intervals) compared to the sub-M$_{\rm Ch}$ results of \cite{bravo19} evaluated at different percentages of shocked (and thus visible) ejecta. Each row of plots is a pair, with labels split between the two plots.}}
\label{fig:B19sub_all}
\end{figure*}

\begin{figure*}
\begin{center}
\textbf{Shen et al. (2018) sub-M$_{\rm Ch}$ D$^6$ Nucleosynthesis from Outer X\% of Ejecta} \\
\includegraphics[width=0.95\textwidth]{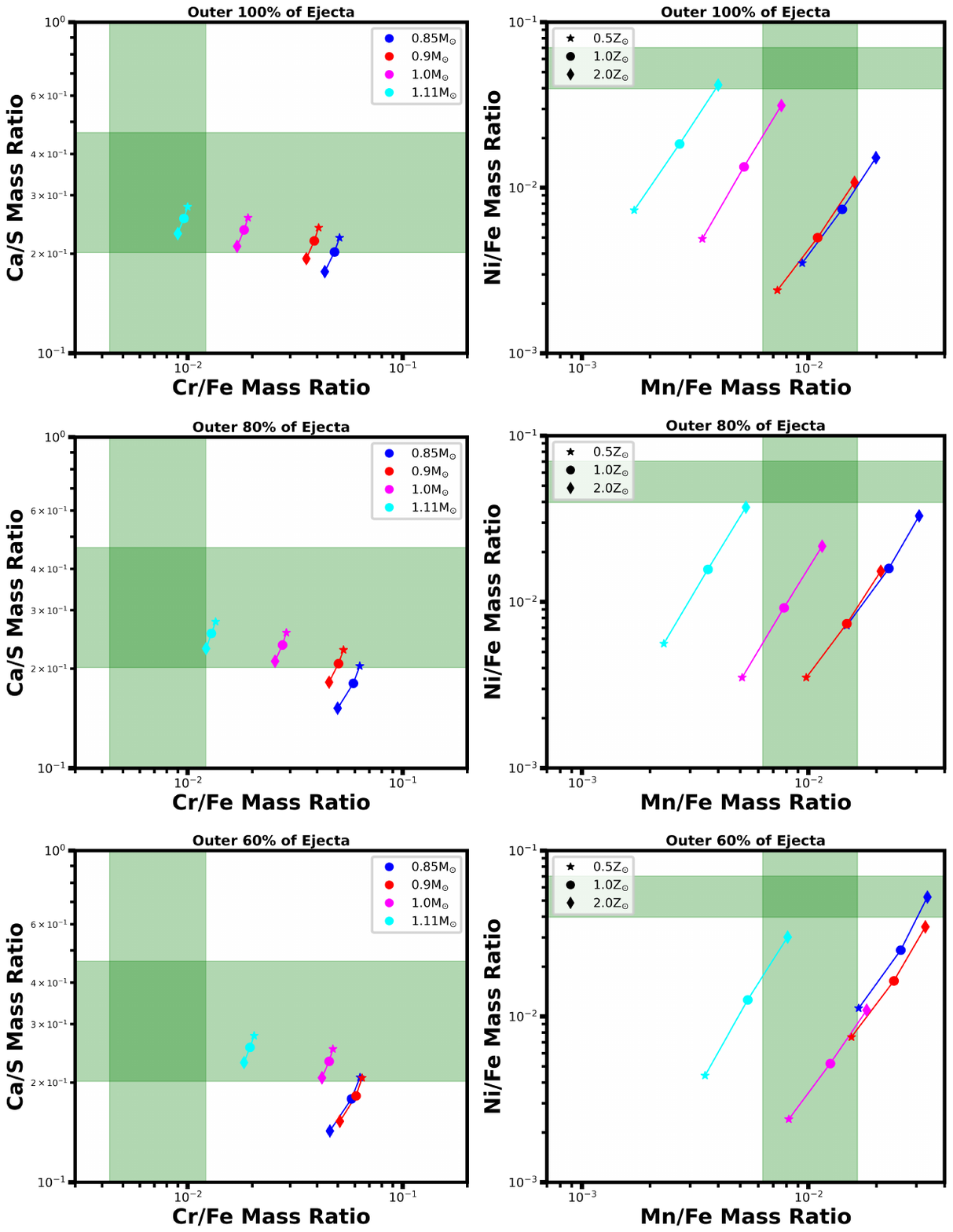}
\end{center}
\vspace{-5mm}
\caption{\footnotesize{Our estimated mass ratios for ejecta in Kepler's SNR (green bars, representing 90\% confidence intervals) compared to the sub-M$_{\rm Ch}$ results of \cite{shen18b} evaluated at different percentages of shocked (and thus visible) ejecta. Each row of plots is a pair, with labels split between the two plots.}}
\label{fig:S18_all}
\end{figure*}

Figures~\ref{fig:S13_all}, \ref{fig:B19_all}, \ref{fig:B19sub_all}, and \ref{fig:S18_all} present detailed results from various Type Ia supernova simulations, showing the estimated ejecta mass ratios for various percentages of reverse-shocked ejecta and initial parameter combinations.

\bibliography{Kepler_MR}

\end{document}